\documentclass[12pt]{iopart}
\usepackage{iopams}
\usepackage{cite}
\expandafter\let\csname equation*\endcsname\relax

\expandafter\let\csname endequation*\endcsname\relax

\usepackage{amsmath}
\usepackage{amsfonts}
\usepackage{amssymb}
\usepackage{color}
\usepackage{epstopdf,cancel,ulem}
\usepackage{epsf,latexsym,bbm,euscript}
\usepackage{times,graphics}
\usepackage{soul,xcolor}
\usepackage{epsfig}

\usepackage{graphicx}
\usepackage{graphicx}% Include figure files
\usepackage{dcolumn}% Align table columns on decimal point
\usepackage{bm}% bold math
\usepackage[colorlinks]{hyperref}% add hypertext capabilities
\usepackage{braket}
\usepackage{color}

\newcommand{\ben}{\begin{eqnarray}}
\newcommand{\een}{\end{eqnarray}}

\newcommand{\be}{\begin{equation}}
\newcommand{\ee}{\end{equation}}
\newcommand{\ba}{\begin{eqnarray}}
\newcommand{\ea}{\end{eqnarray}}

\newcommand{\beq}{\begin{equation}}
\newcommand{\eeq}{\end{equation}}
\newcommand{\beqa}{\begin{eqnarray}}
\newcommand{\eeqa}{\end{eqnarray}}

\def\bwt{\begin{widetext}}
\def\ewt{\end{widetext}}

\newcommand {\h}[1]{\hat#1{}}

\newcommand{\mean}[1]{ \left \langle #1 \right \rangle}

\begin{document}

\title[Emergent Dark Energy in Classical Channel Gravity with Matter ]{Emergent Dark Energy in Classical Channel Gravity with Matter }

  \author{Romain Pascalie$^{1,2}$,Natacha Altamirano$^{1,3*}$, Robert B. Mann${^1}$}
  \address{
$^{1}$Department of Physics and Astronomy, University of Waterloo, Waterloo, Ontario, Canada, N2L 3G1. \\
$^{2}$Department of Physics, \'Ecole Normale Sup\'erieure de Lyon, Lyon Cedex 07, France. \\
$^{3}$Perimeter Institute, 31 Caroline St. N. Waterloo
 Ontario, N2L 2Y5, Canada. \\
}
\ead{$^*$naltamirano@perimeterinstitute.ca}

\vspace{10pt}
\begin{indented}
\item[]\today
\end{indented}

\begin{abstract}
Motivated by the recent increased interest in energy non-conserving models in cosmology, we extend the analysis of the cosmological consequences of the Classical Channel Model of Gravity (CCG). This model is based on the classical-quantum interaction between a test particle and a metric (classical) and results in a theory with a modified Wheeler-deWitt equation that in turn leads to
non conservation of energy. We show that CCG applied to a cosmological scenario with primordial matter leads to an emergent dark fluid that at late times behaves as a curvature term  in the Friedmann equations, showing that the late time behaviour is always dominated by the vacuum solution. We  discuss possible observational constraints for this model and that  -- in its current formulation  -- CCG eludes any meaningful constraints from current observations.
\end{abstract}

%\maketitle

\section{\label{sec:level1}Introduction}

Much effort has devoted over several decades to unify quantum mechanics (QM) with general relativity (GR). These two theories have been very successful in their own domains, but as yet are not compatible with each other.  Even though the universe on large scales is well described by the classical Einstein equation there are still astrophysical and cosmological regimes that evidently are in need of a quantum theory of gravity (at least in principle) to be properly explained: the primordial singularity, inflation, black hole physics, among others. The available techniques to study these problems (quantum field theory in curved space time), while providing some insight, leave open questions that require  a deeper understanding of the interplay between QM and GR. Notwithstanding the existence of well known problems in consistently combining quantum and classical dynamics \cite{PhysRevA.63.022101}, some authors have questioned if gravity actually needs to be quantized \cite{carlip_is_2008,albers_measurement_2008}. 

Recently attention has been drawn to the Classical Channel model of Gravity (CCG), which  proposes that gravity is purely classical, in the sense of being unable to transmit quantum information \cite{2014NJPh...16f5020K}. This model is based on the fact that any gravitational source cannot be shielded; it is thus possible to always gain partial information of a source via weak measurements \cite{aharonov_how_1988}. CCG has been shown to yield similar consequences to models of gravitational decoherence by Diosi and Penrose \cite{diosi_models_1989,penrose_gravitys_1996}, despite their fundamental differences \cite{Altamirano:2016fas}. In  CCG two quantum systems interact gravitationally via a sequence of weak measurements and feedback that {\it{simulate}} the gravitational potential. The overall dynamics is then given by a non-unitary Lindblad  master equation, whose systematic (unitary) term contains the gravitational potential and a decoherence term describing the noise in the interaction, ensuring a classical information exchange that forbids entanglement. 

%The Classical Channel model of Gravity 
CCG can be also be understood as model that enables quantum-classical interactions \cite{Altamirano:2016hug} in the spirt of \cite{diosi_quantum_2000,diosi_quantum_1995}. In this context, two quantum systems interacting gravitationally will generate a potential that is by assumption classical. It was shown, that in order to maintain a density matrix with appropriate physical properties 
the   noise should be added in the quantum dynamics (and conversely in the classical potential). This noise term introduces  decoherence leading to a break in unitary dynamics. Non-unitary dynamics then appears in general  in the evolution of quantum systems interacting with an environment.  Quantum-classical interactions inducing non unitary dynamics have been studied in \cite{Yang:2017xyh}, and also in the particular case of gravitational induced decoherence  \cite{1742-6596-306-1-012006,PhysRevD.93.024026} . On the other hand, non unitary dynamics and fundamental decoherence have been found to govern dynamics emergent from quantum gravity \cite{PhysRevLett.80.2508,PhysRevLett.93.240401} and yielded light in the resolution of the black hole information paradox \cite{PhysRevD.14.2460,PhysRevD.52.2176}.  Finally, decoherence and non unitary dynamics have been used to study emergent unitary evolution in quantum systems \cite{Altamirano:2016knc,2016arXiv160504302G};  these concepts are at the core of collapse models \cite{bassi_models_2013} and could imply interesting observable consequences for the universe \cite{Josset:2016vrq}. Some consequences of the Newtonian CCG model have been the subject of a number of recent investigations \cite{2014NJPh...16f5020K,Kafri:2015iha,Khosla:2016tss,Altamirano:2016fas}.

Recently \cite{Altamirano:2016hug}, it has been shown that the CCG model applied to the cosmological scenario leads to an emergent dark fluid as a consequence of the non conservation of energy given by the modification of quantum dynamics. In this context, CCG postulates that the metric is sourced by a quantum {\it{scale factor}} and its conjugate momentum whilst a quantum test particle will move according to geodesics given by a classical emergent metric. An observer  unaware of the quantum nature of the underlying degrees of freedom will describe the universe by a metric given by a modification of the Wheeler-deWitt equation.

The analysis in \cite{Altamirano:2016hug} was restricted to the study of CCG in the context of an empty universe with the aim of understanding in depth the emergent dark fluid. However if CCG is to be regarded as the fundamental description of classical gravity  it is then also fundamental in describing the interplay between gravity and matter. This paper is the first attempt in investigating this aspect of CCG: we consider a universe  filled with dust, which serves as  the effective stress-energy of some primordial weakly interacting matter. Even though dust is not the most general matter source, this paper sheds light on the interplay of matter with a universe dominated by quantum variables that  is still fully classical for any observer. For the analysis, we choose to adopt a model in which all the noise is in the scale factor (similar to the description presented in \cite{Altamirano:2016hug});  the primordial dust acts as a {\it{spectator}}, modifying the initial Hamiltonian governing the equations of motion of the scale factor. This is not the only option nor is it the most general, and we will discuss other possibilities for how to treat the primordial dust in the concluding section of our paper.

In this paper we analyze the behaviour of both  the emergent fluid and the primordial perfect fluid focusing on the differences with respect to the empty universe model \cite{Altamirano:2016hug}. Extending  this study, we provide an improved  description of the emergent dark fluid and find that it can be characterized as emergent curvature in the universe.

Our paper is organized as follows.  In Sec.~\ref{Model} we introduce the basic notions of the CCG model and how to couple it with a dust fluid.  In Sec.~\ref{seceq} we derive the equations of motion for our system and give a discussion about the conservation of the Hamiltonian. Section \ref{secresults} is devoted to the results, where we present the behaviour for the scale factor, the emergent fluid and the behaviour and consequences of the primordial dust. We conclude this work in sections \ref{secdis} and \ref{secconc}, giving a discussion about the observable consequences of CCG and future directions. \ref{app1} and \ref{ahamiltonian} contain the relevant notation for our work, a derivation of the Hamiltonian and the relation with quantities used in this work with the physical observable quantities.

\section{Model and strategy} \label{Model}

We consider an empty, homogeneous and isotropic universe described by the FRW metric:
\begin{equation}
   ds^2 = -N^2dt_\texttt{d}^2 + a_p^2(t_\texttt{d})\left( \frac{dr^2}{1-kr^2}   + r^2  d\Omega \right) \,,
\end{equation}
where $N$ is the lapse function, $a_p(t_\texttt{d})$ is the scale factor, $k$ is the spatial curvature, and $t_\texttt{d}$ is the proper time associated with a fluid filling the universe. The action for a spacetime filled with a pressureless perfect fluid (or dust) is:
\begin{equation}
      \mathcal{S} =\int d^4x \, \sqrt{-\mathrm{g}} \left( \frac{1}{2\kappa}\mathrm{R} - \rho \right)\,, \label{action2}
\end{equation} 
where $\mathrm{g}$ is the determinant of the metric $\mathrm{g}_{\mu\nu}$, $\mathrm{R}$ is the Ricci scalar, $\rho$ the energy density of the fluid and $\kappa=8\pi G c^{-4}$. The dust fluid is characterized by its energy momentum tensor
\beq
T_\texttt{d}^{ab}=\rho\, U^a_\texttt{d}U^b_\texttt{d}\,,\,\,\,\,\,\,\,\,\, U^a_\texttt{d}=\bigg(\frac{\partial}{\partial t_\texttt{d}}\bigg)^a\,,
\label{dustT}
\eeq
where $U^a_\texttt{d}$ is the four velocity of the dust. The dynamics of the fluid are determined by the conservation of the stress-energy
\beq
\nabla_a\,T_\texttt{d}^{ab}=0\,.
\label{dustcons}
\eeq
Eqs.\eqref{dustT} and \eqref{dustcons} allow as to write the energy density as a function of the scale factor
\beq
\rho(t_\texttt{d})=\frac{\rho_{0p}}{a_p(t_\texttt{d})^3}\,,
\label{dust_scale}
\eeq
where $\rho_{0p}$ is the energy density at a reference time $t_0$ such that $a_p(t_0)=1$\,. In \ref{ahamiltonian}  we find that the Hamiltonian associated with \eqref{action2} is
%
%Using the conservation of energy we know that for dust $\rho(\tau) = \frac{\rho(0)}{a^3(\tau)}$. Hence, from the action we found the following Lagrangian:
%\begin{equation}
%  \mathcal{L} = ka^2N - \frac{a'^2}{N} - N\rho_0 a
%\end{equation}
%where $a' =\frac{da}{d\tau}$ and we have defined $\rho_0 = \frac{8p}{3}\rho(0)$.
%The conjugate momentum of the scale factor is $p = - \frac{2a'}{N}$, we then have the Hamiltonian density:
\begin{equation}
  H= N\left( -\frac{p^2}{4} - ka^2 +\rho_0 a \right)\,,
  \label{hamil}
\end{equation}
where $a$ and $p$ are associated with the physical scale factor $a_p$ and its conjugate momentum by Eqs.~\eqref{relationscale} and we have set $c=1$.
 
The standard procedure in quantum cosmology is to canonically quantize the degrees of freedom $a$ and $p$ and solve the Wheeler-deWitt equation $\hat{H}\Psi(\h{a},\rho)=0$ for the wave function of the universe $\Psi(\h{a},\rho)$. This process is the quantum analog of imposing the classical Hamiltonian constraint $\mathcal{H}=0$, which indeed reduces to the first Friedman equation. Note that the Wheeler-deWitt equation can be reformulated in terms of the density matrix $\varrho$ associated with the wave function $\Psi$ as
\beq
  \frac{d\hat{\varrho}}{d\tau}=-\frac{i}{\hbar}[\hat{H},\varrho]\,.
\label{motion}
\eeq

As described in detail  in \cite{Altamirano:2016hug},   CCG applied to cosmology induces a noise term in Eq.\eqref{motion} yielding a master equation
\begin{equation}
  \frac{d\hat{\varrho}}{d\tau} = -\frac{i}{\hbar} [\hat{H}, \hat{\varrho} ] - \frac{\gamma}{8} [\hat{a}^2,[\hat{a}^2, \hat{\varrho} ] ] \,,\label{master2}
\end{equation}
that fully describes the dynamics. 
 The second term in the master equation characterizes a quantum noise experienced by the scale factor due to its continuous weak interactions with a bath of hidden particles, and it is controlled by a strength parameter $\gamma$ (a full derivation of these equations for cosmology was first presented in \cite{Altamirano:2016hug} and details concerning the relation between $\gamma$ and the decoherence parameter can be found in ~\ref{amaster}). 
 
In spirit, CCG changes ordinary unitary quantum dynamics by adding a decoherence term. As discussed in the introduction, such terms are generic for the evolution of open systems and for theories considering classical-quantum interactions. CCG is strongly based on the premise that any gravitational system is necessarily an open system and continuously interacting via weak quantum measurements due to the impossibility of gravitational shielding \cite{aharonov_how_1988,PhysRevA.36.5543}.  The first investigation of 
CCG in the context of quantum cosmology indicated  that the effect of  \eqref{master2}  is to  
induce a perfect fluid for classical observers unaware of quantum noise \cite{Altamirano:2016hug}.  To understand this statement better, note that  any classical observer will be unaware of the underlying quantum noise in the scale factor. They will therefore describe the dynamics of the universe following the Einstein equations 
 \beq
 G^{ab}=R^{ab}-\frac{1}{2}R\,g^{ab}= \kappa T^{ab}\,,
 \label{EE}
 \eeq 
  where $T^{ab}$ is the  classical energy-momentum tensor inferred from  measurements of the behaviour of
  the classical scale factor $\texttt{a}_p$ by this observer, who describes spacetime using the FRW metric  
  \beq
  ds^2_c=\texttt{a}_p^2(\tau)\left( -d\tau^2+\frac{dr^2}{1-kr^2}   + r^2  d\Omega_2 \right) \,,
  \label{new_metric}
  \eeq
 since CCG does not modify either the conformal coordinates or the symmetries of the spacetime. 
The observed scale factor is computed using

  \beq
  \texttt{a}_p^2=\mean{\hat{a}_p^2}=\text{Tr}(\hat{a}_p^2 \hat{\varrho})\,,
  \label{classicala_def}
  \eeq
upon solving the master equation \eqref{master2}.
Note that if the Wheeler-deWitt equation \eqref{motion} was instead used, we would obtain from \eqref{classicala_def}   the Friedmann equations with the energy density defined in \eqref{dust_scale}. The change in quantum dynamics due to \eqref{master2} implies we should expect to obtain  different behaviour for   $\texttt{a}_p$ than would otherwise be obtained
using  \eqref{motion}. A classical observer, necessarily using the Einstein Equations \eqref{EE},  will still obtain the Friedman Equations but with a different stress-energy  $T^{ab}$. The conservation of $T^{ab}$ and Eq.~\eqref{classicala_def} are enough to fully characterize the resultant classical perfect fluid. As discussed in ~\ref{obsphys} the scale factor $a_p$ is related to $a$ via $a=a_p\sqrt{3V_0/\kappa}$ and thus the observed scale factor $\texttt{a}_p$ is related to the quantities involved in Eq.~\eqref{master2} as
$\texttt{a}=\texttt{a}_p\sqrt{3V_0/\kappa}$, where $V_0$ is a fiducial volume as discussed in \ref{ahamiltonian}.
Our plots will show quantities related to $\texttt{a}$ and when necessary we will comment on the relation with the observed scale factor $\texttt{a}_p$.

    \section{Equations of motion}\label{seceq}
    
    In order to find the equations of motion we make use of the formula $\mean{\h A}= \mathrm{Tr}(\hat{A}   {\hat{\varrho}})$ together with the master equation \eqref{master2} to find:
  \begin{align}
     \frac{d\left<\hat{a}\right>}{d\tau} &= - \frac{\mean{\hat{p}}}{2}c^2\,, \label{eq1} \\
     \frac{d\left<\hat{p}\right>}{d\tau} &=  2k\left<\hat{a}\right> - \rho_0 \,,  \label{eq2}\\
     \frac{d\left<\hat{a}^2\right>}{d\tau} &= - \frac{\left<\hat{p}\hat{a}+\hat{a}\hat{p}\right>}{2}c^2\,, \label{sys} \\
     \frac{d\left<\hat{p}^2\right>}{d\tau} &= 2k \left<\hat{p}\hat{a}+\hat{a}\hat{p}\right> + \gamma\, \hbar^2 \left<\hat{a}^2\right> - 2\rho_0\left<\hat{p}\right> \label{eq4}\\
     \frac{d\left<\hat{p}\hat{a}+\hat{a}\hat{p}\right>}{d\tau} &= - c^2\left<\hat{p}^2\right> + 4k\left<\hat{a}^2\right>  - 2\rho_0 \left<\hat{a}\right>\,. \label{eq5}
\end{align}
These five equations form a closed coupled system of differential equations. By comparing with the equations of motion found in \cite{Altamirano:2016hug} we note that the presence of dust in the background  couples Eqs.\eqref{sys}-\eqref{eq5} with Eqs.\eqref{eq1}-\eqref{eq2}. Nevertheless, Eqs.\eqref{sys}-\eqref{eq5} decouple from each other at third order:

\begin{align}
    \left<\hat{a}^2\right>^{'''}+ 4k c^2\left<\hat{a}^2\right>^{'} -\frac{\gamma \hbar^2 c^4}{2}\left<\hat{a}^2\right> &= -\frac{3}{2}c^4\rho_0\left<\hat{p}\right>\,, \label{eqc1}\\
    \left<\hat{p}^2\right>^{'''} + 4k c^2\left<\hat{p}^2\right>^{'} -\frac{\gamma\hbar^2 c^4}{2}\left<\hat{p}^2\right> &= \rho_0 \gamma\,\hbar^2 c^2\left<\hat{a}\right>\,, \label{eqc2}\\
     \left<\h \Gamma\right>^{'''} + 4k c^2\left<\h \Gamma\right>^{'} -\frac{\gamma\hbar^2 c^4}{2}\left<\h \Gamma\right> &= 3 c^2  \rho_0(2 k \left<\hat{a}\right> - \rho_0)\,, \label{eqc3}
\end{align}
where we have renamed $\h a\h p+\h p\h a=\h \Gamma$ and make use of the notation $d/d\tau=\,'$.

In the next section we shall present the solutions to these equations and analyze the various resulting  behaviours. However, before doing so, we first  analyze the behaviour of the Hamiltonian operator in our model. From the system of equations \eqref{eq1}-\eqref{eq5} we find that 
\begin{equation}\label{emergent}
       \frac{d\left<\hat{H}\right>}{d\tau} = -\frac{\gamma}{4}(c\hbar)^2\left<\hat{a}^2\right>\,,
\end{equation}  
which is only equal to zero if $\gamma=0$. Note that the Wheeler-DeWitt equation will indeed conserve the quantity $\left<\hat{H}\right>$ since the Hamiltonian operator commutes with itself at all times. If $\left<\hat{H}\right>$ is regarded as a measure of the classical energy emergent from a quantum theory of gravity, we see that CCG predicts the existence of additional energy. In this sense it is another example of  energy non-conservation theories as unimodular gravity (see \cite{Padilla:2014yea} and references therein)  that might present interesting observable consequences \cite{Josset:2016vrq,Altamirano:2016hug}.  We emphasize that any observer unaware of the quantum nature of spacetime and 
describing its dynamics via the Einstein equations will see this extra energy manifest as a dark energy fluid.  
%One can therefore understand our model in two different ways: either as a non-conserving energy theory or as a modify theory of gravity -- any modification of gravity can be understood as Einstein theory plus an extra fluid []. 

\section{Results}\label{secresults}

In this section we present the result of our analysis, in particular for the scale factor and the emergent dark energy fluid due to the modification of the Wheeler-DeWitt equation \eqref{master2}. To this end we are going to solve for the scale factor $\texttt{a}=\sqrt{\mean{a^2}}$ in conformal time $\tau$ using the coupled system of equations \eqref{eqc1}-\eqref{eqc2} and convert to proper time $t$ using the relation $\texttt{a}(\tau)d\tau=dt$. Note that for  $\rho_0=0$ the solution for the scale factor can be written as
\cite{Altamirano:2016hug}
\beq
\texttt{a}^2=\sum_{i=1}^3A_i\exp (\omega_i\tau)\,,
\label{sol_homo}
\eeq
where the coefficients $A_i$ depend on the initial conditions, $k$ and $\gamma$, and where the characteristic equation
\beq
\omega_i^3+4kc^2\omega_i-\frac{\gamma \hbar^2 c^4}{2}=0 
\label{charac_eq}
\eeq
yields the solution for $\omega_i$.

In our case, because the system of equations is inhomogeneous, our solution for the scale factor will be the homogenous solution plus a particular solution of \eqref{eqc1}
\begin{align}
&  \texttt{a}^2=\sum_{i}^3A_i^{(k)}\exp (\omega_i\tau) \nonumber\\
&\qquad  +\left\{
  \begin{array}{@{}ll@{}}
   \quad  -\frac{3}{\gamma}\rho_0^2\tau+3\frac{{p_0}\rho_0}{\gamma}, & k=0 \\
    3\tilde{A}\cos(\sqrt{k}\tau)-\tilde{B}\frac{3}{\sqrt{k}}\sin(\sqrt{k}\tau), & k=\pm 1
  \end{array}\right.
  \label{scale_sol}
\end{align} 
where we have used units where $c=\hbar=1$ and 
\begin{eqnarray}
\tilde{A}=\frac{\rho_0}{36 k^3+\gamma^2}(12 k^2 a_0+\gamma {p_0}-6k \rho_0)\,, \\ 
\tilde{B}=\frac{\rho_0}{36 k^3+\gamma^2}(6 k^2{p_0}-2 a_0k \gamma+\gamma \rho_0)\,.
\end{eqnarray}
Here $A_i^{(k)}=A_i^{(k)}[k,\gamma,a_0,{p_0},\rho_0,a_{20},{p_{20}},\Gamma_0]$ is a function of the physical parameters and initial conditions, where $a_0=\mean{\h a}(0)$, ${p_0}=\mean{\h {p}}(0)$, $a_{20}=\mean{\h a^2}(0)$, ${p_{20}}=\mean{\h {p}^2}(0)$ and $\Gamma_0=\mean{\h \Gamma}(0)$. The initial conditions must be chosen in such a way that the uncertainty principle holds
\beq\label{uprin}
\bigg(\mean{\h a^2}-\mean{\h a}^2\bigg)\bigg(\mean{\h {p}^2}-\mean{\h {p}}^2\bigg)\geq \bigg(\frac{1}{2}\mean{\h \Gamma}-\mean{\h a}\mean{\h {p}}\bigg)^2+\frac{1}{4}\,.
\eeq

In the sequel we choose to work with $a_0={p_0}=\gamma_0=0$,  ${p_{20}}=1$ and choose $a_{20}=1/4$ in order to saturate the uncertainty principle. This states are called coherent states with minimum uncertainty.
 
\subsection{Scale Factor}

\begin{figure*}
\minipage{0.32\textwidth}
  \includegraphics[width=\linewidth]{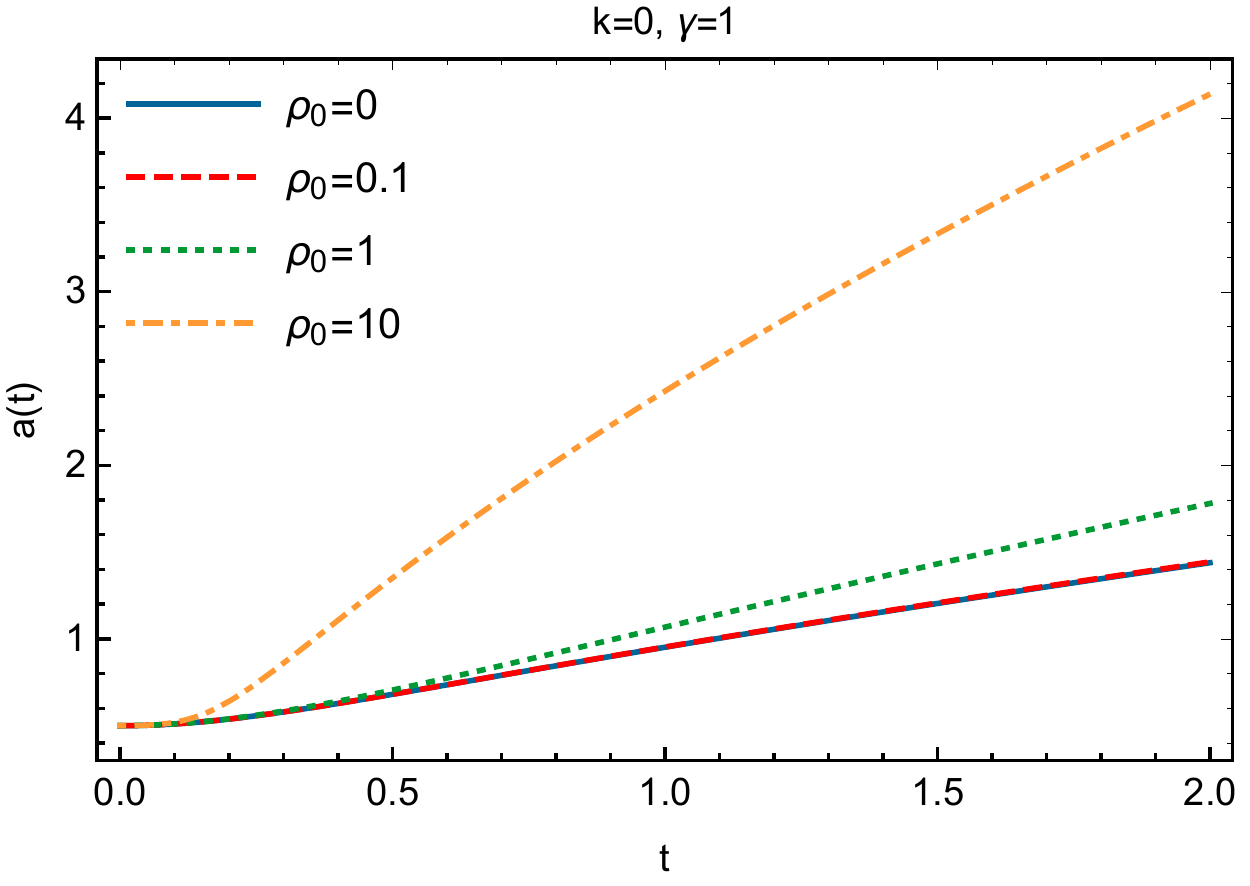}
  \endminipage\hfill
\minipage{0.32\textwidth}
  \includegraphics[width=\linewidth]{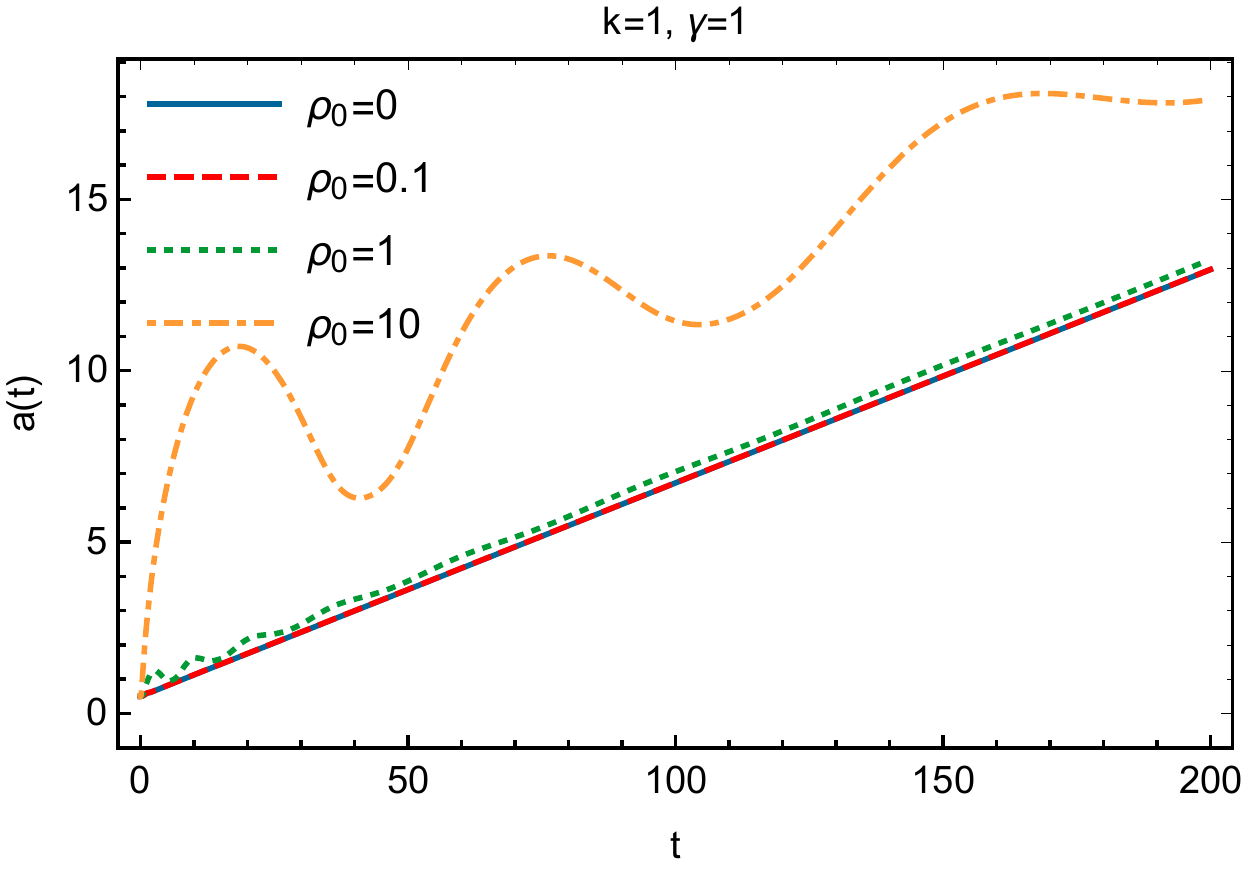}
  \endminipage\hfill
\minipage{0.32\textwidth}
  \includegraphics[width=\linewidth]{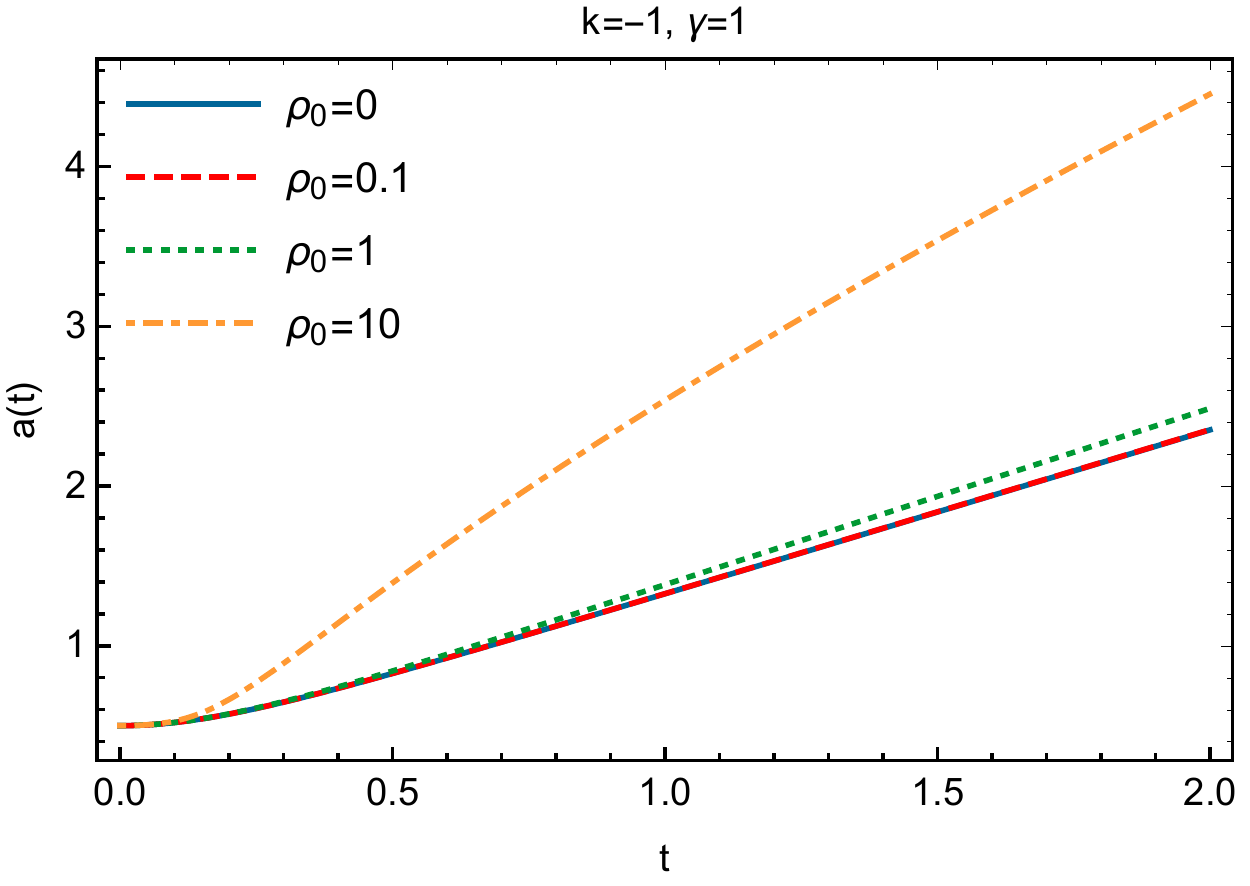}
  \endminipage\\
\minipage{0.32\textwidth}
  \includegraphics[width=\linewidth]{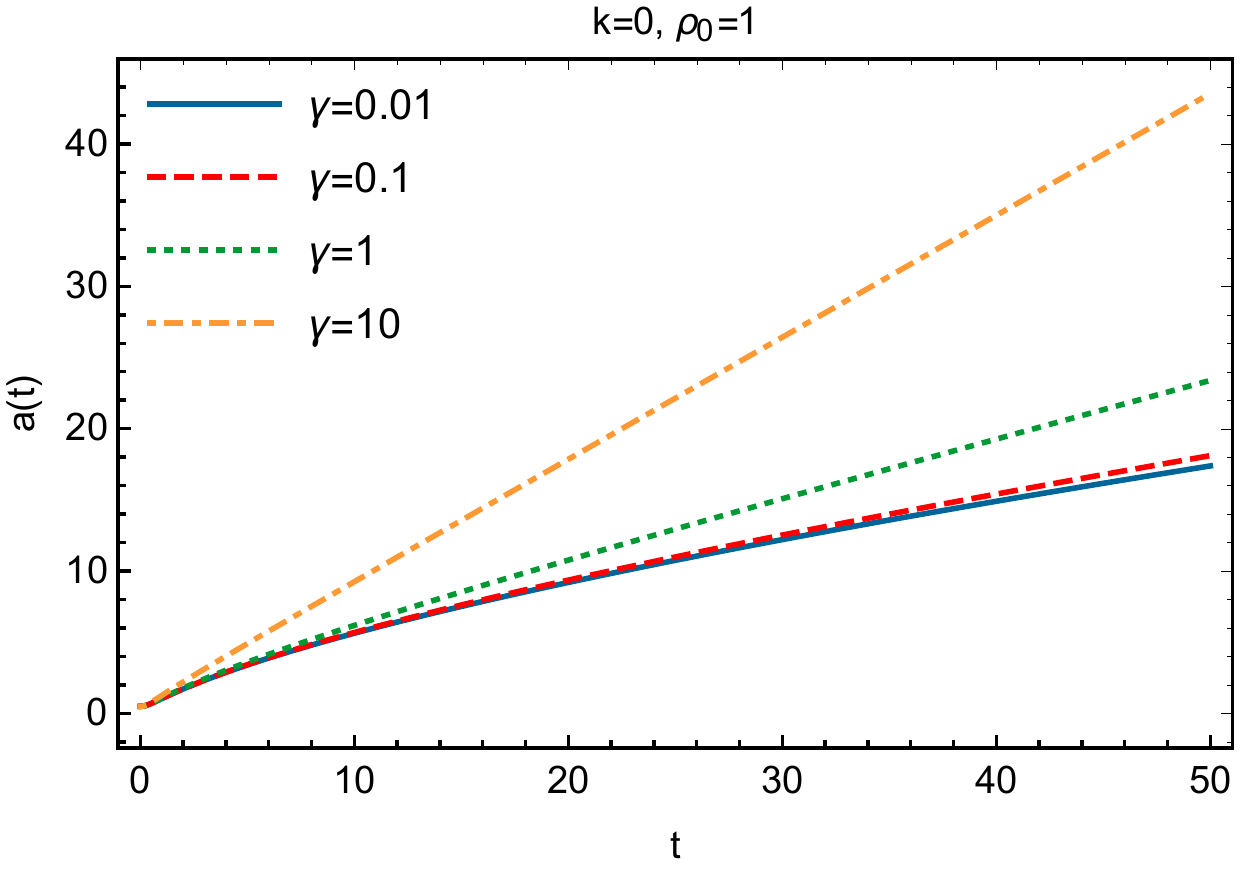}
  \endminipage\hfill
\minipage{0.32\textwidth}
  \includegraphics[width=\linewidth]{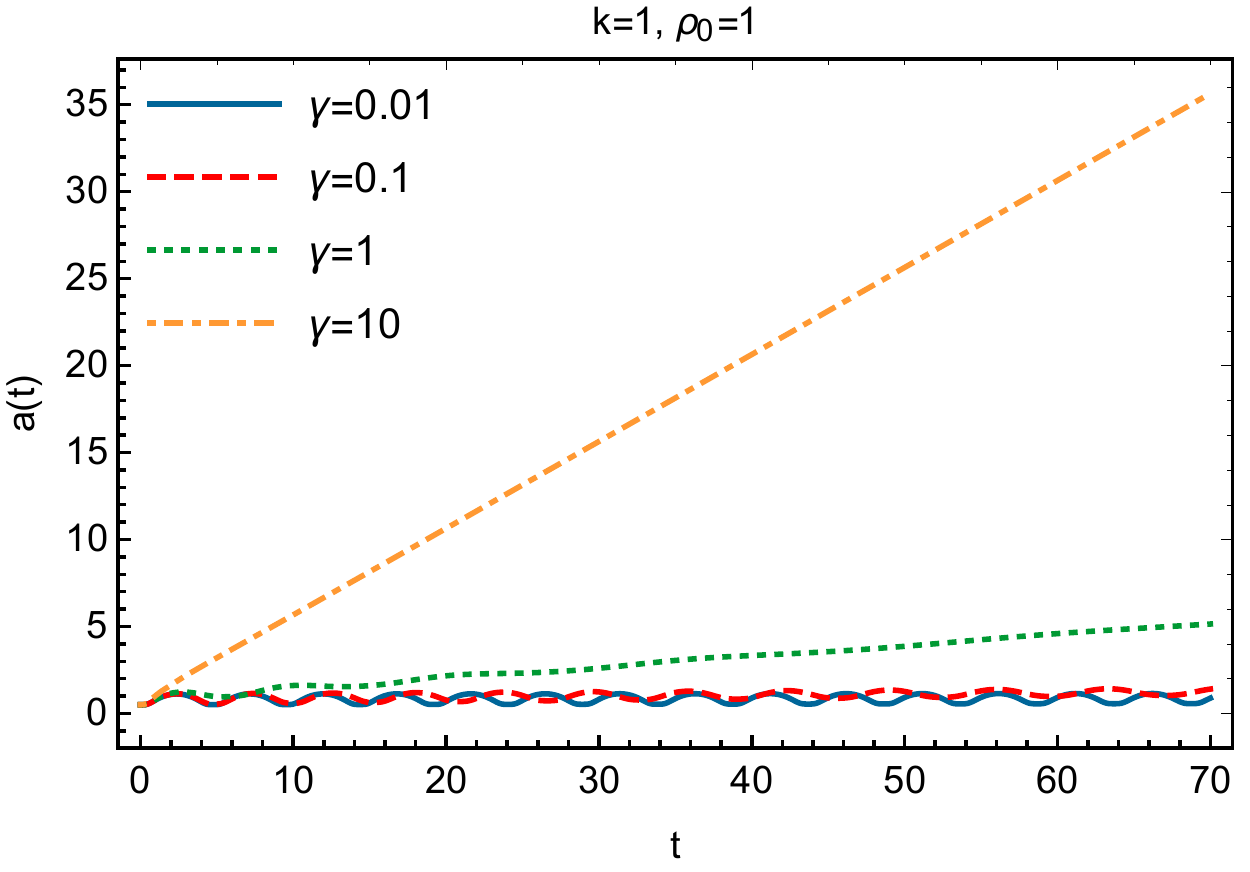}
  \endminipage\hfill
\minipage{0.32\textwidth}
  \includegraphics[width=\linewidth]{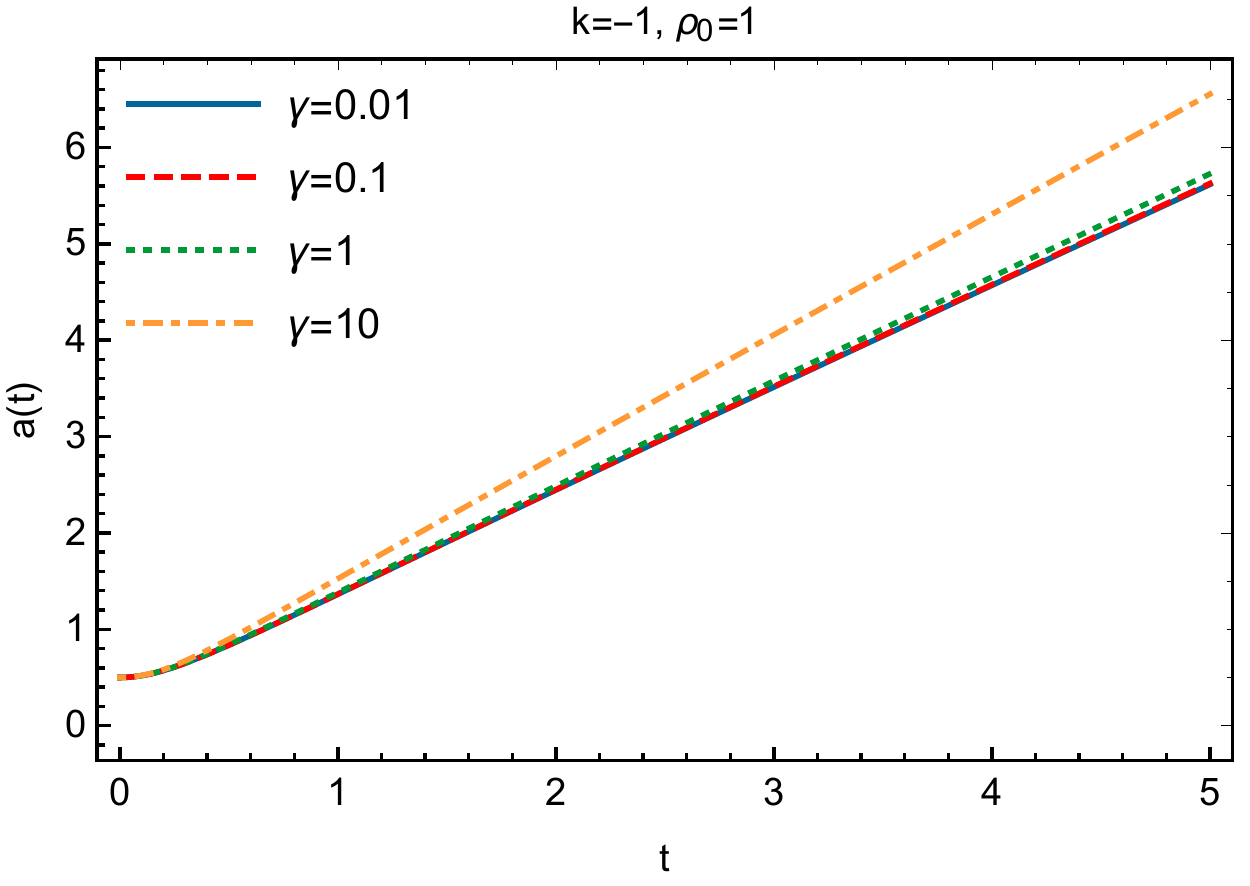}
  \endminipage
  \caption{\label{fig:a0}  Evolution of the scale factor in comoving time for $k=0$ (left), $k=1$ (middle ) and $k=-1$ (right). We show results for coherent states with minimal uncertainty, varying $\rho_0$ (top) and $\gamma$ (bottom). In all cases we find that the late time behaviour asymptotes to linear growth;  for $k=1$ the universe is oscillatory at early times 
  that are dominated by small $\gamma$ and/or large $\rho_0$.}
 \label{fig:scale}
\end{figure*}

In this section we present the results for the behaviour of the scale factor under the classical channel gravity when varying $\gamma$ and $\rho_0$ for the three different values of $k$. In figure~\ref{fig:scale}
we depict the scale factor $\texttt{a}$ as a function of proper time for fixed $\gamma$ (first row) and $\rho_0$ (second row).

\begin{figure*}
\minipage{0.32\textwidth}
  \includegraphics[width=\linewidth]{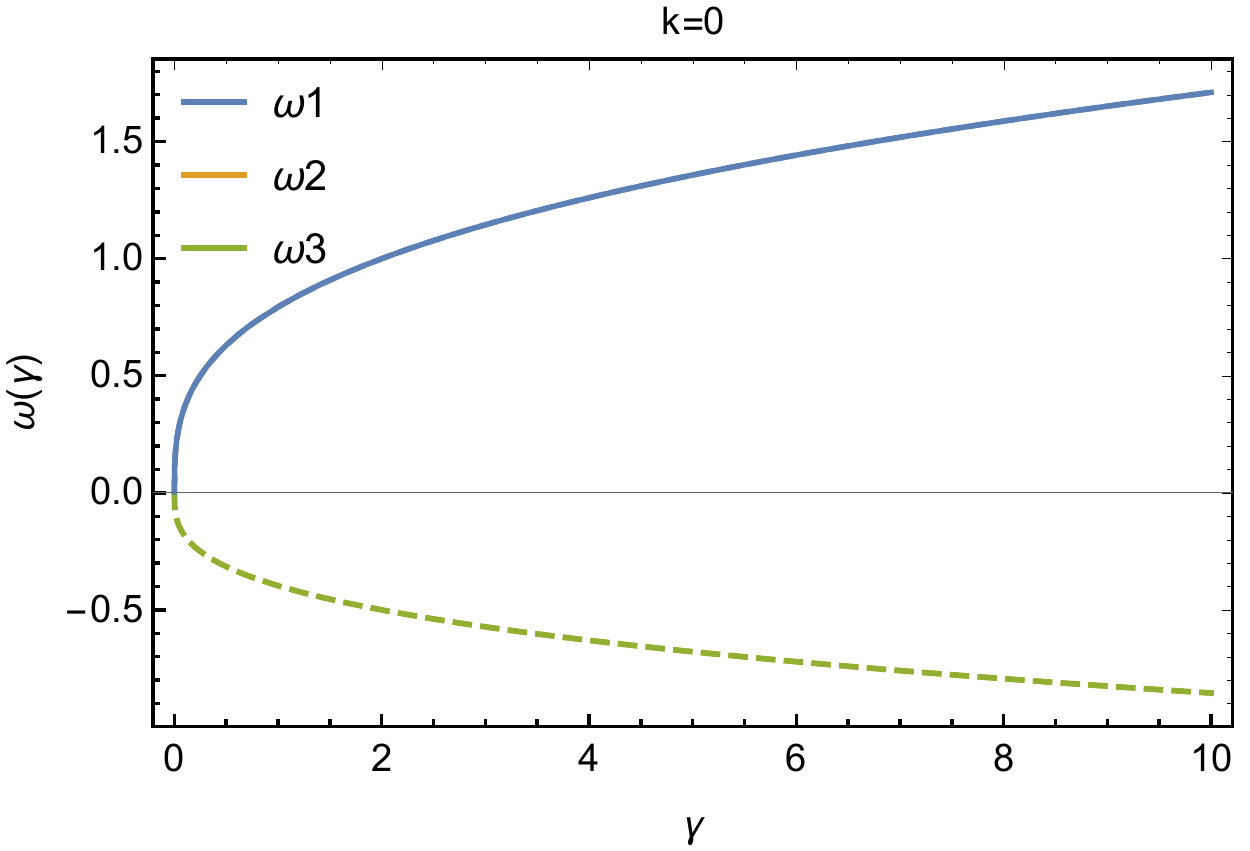}
  \endminipage\hfill
\minipage{0.32\textwidth}
  \includegraphics[width=\linewidth]{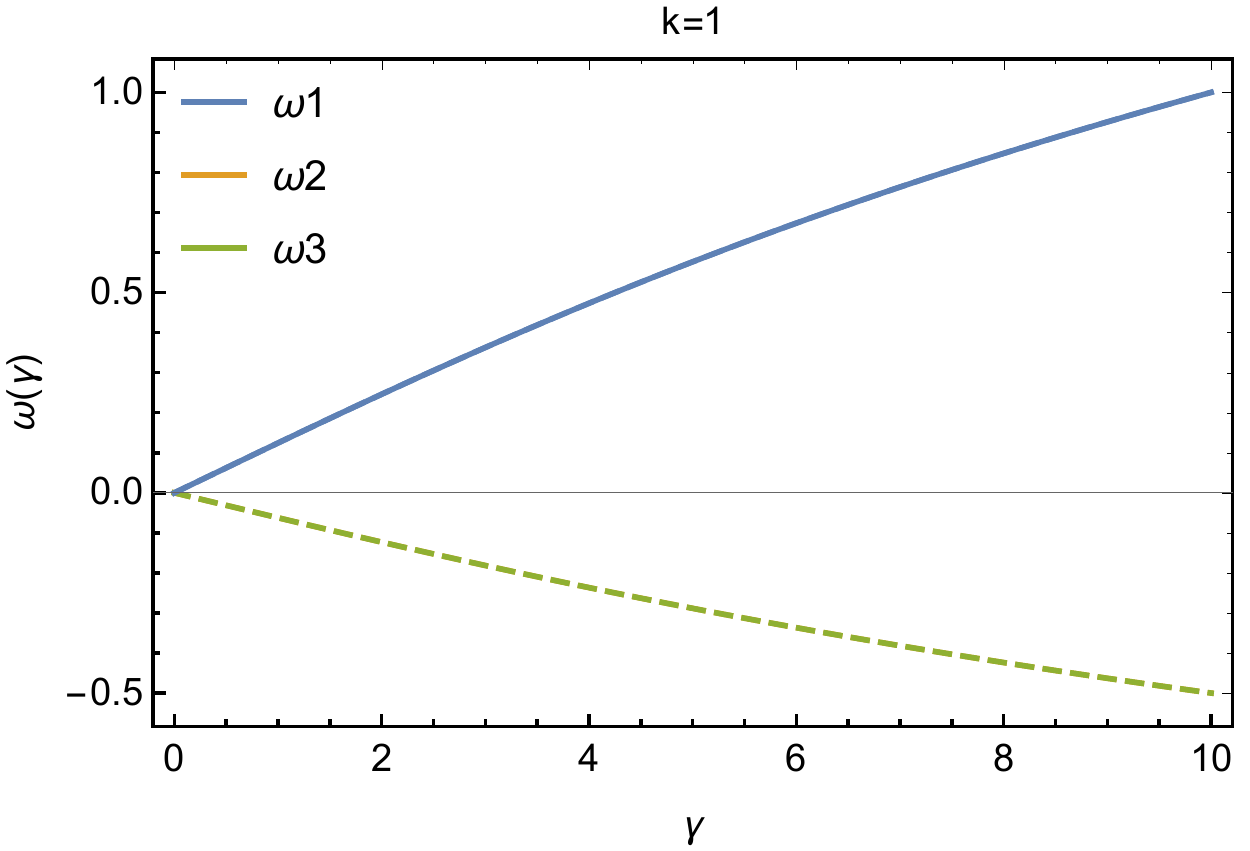}
  \endminipage\hfill
\minipage{0.32\textwidth}
  \includegraphics[width=\linewidth]{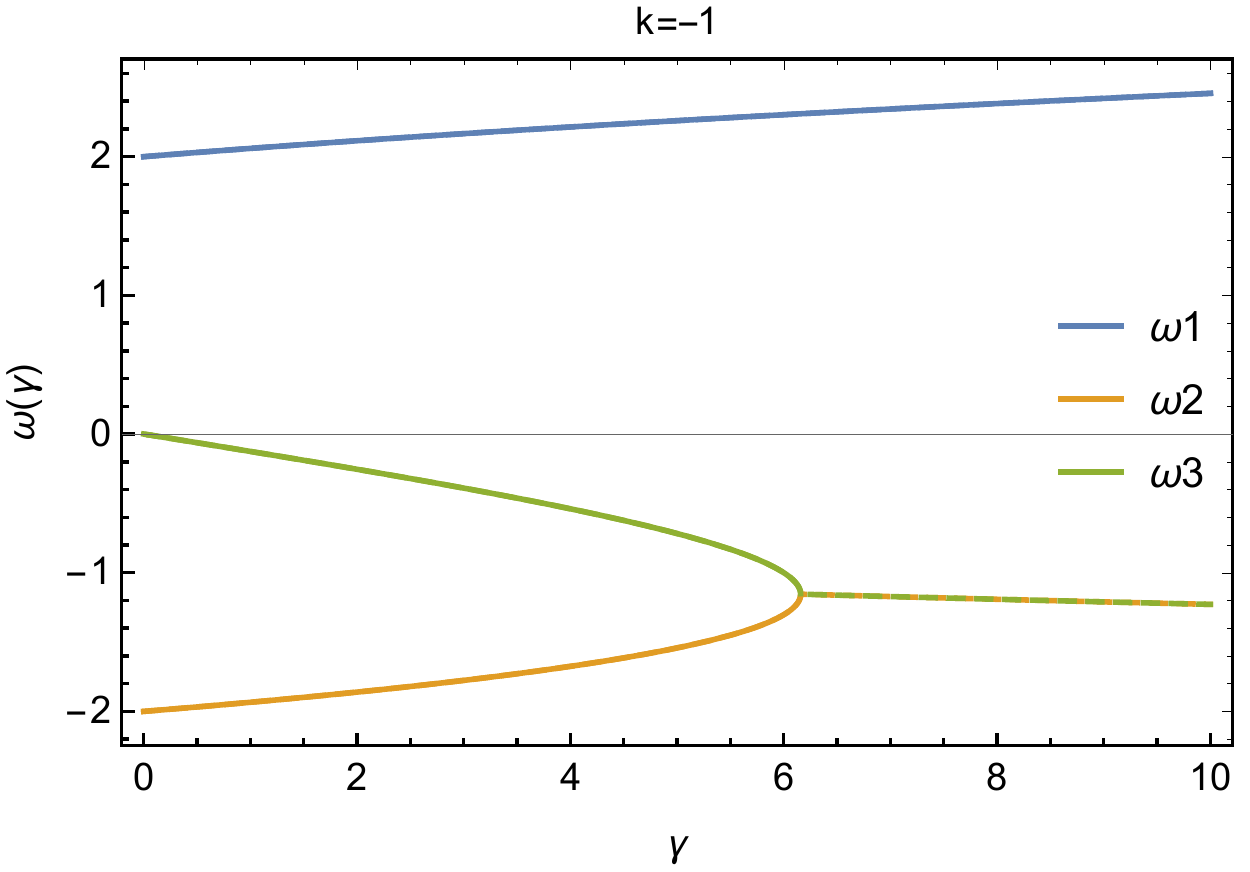}
  \endminipage
  \caption{Behaviour of the roots of the characteristic equation \eqref{charac_eq} for the three different values of $k$. For all cases the characteristic equation admits a positive real root (solid, blue line). In dashed lines we show the real part of the imaginary solutions and for the cases $k=0$ and $k=1$ it coincides for both roots $\omega_1$ and $\omega_2$ (dashed green line -- Note that the orange dashed curve is superposed). For the case $k=-1$ the characteristic equation admits two negative real solutions for $\gamma <6$. The fact that there always exists a positive real solution means that for late time the scale factor will be dominated by this solution.  }
  \label{fig:frequencies} 
\end{figure*}

We find for   $k=1$ that the early time behaviour of the scale factor is dominated by oscillations that become sharper for small $\gamma$ or large $\rho_0$.  At late times, the scale factor behaves linearly in $t$ in all cases, suggesting an exponential expansion as a function of conformal time. The characteristic equation \eqref{charac_eq} always admits a positive real root for any positive value of $\gamma$ and  indeed the solution associated with this root dominates the homogeneous solution. The behaviour of the three different frequencies as a function of gamma is shown in Fig~\ref{fig:frequencies}. Showing in solid lines the pure real root and in dashed the real part of complex roots.

In addition, for $k=0$ and $k=1$ the inhomogeneous solution is subdominant at late times since it grows either linearly or oscillates respectively. The case $k=-1$ is different: the inhomogeneous solution is exponential in conformal time $\sim \exp{\tau}$. This inhomogeneous solution competes with the exponential growth of the homogenous part. However, the rightmost diagram in  Fig.~\ref{fig:frequencies} shows that the real root for $k=-1$ is always bigger than 2, which implies an exponential growth of at least $\sim \exp{2 \tau}$. This analysis shows that for all curvatures the homogenous part of the solution for the scale factor dominates the late time behaviour and thus the late time universe predicted by the classical channel model with matter will behave comparably to the previously studied empty 
case  \cite{Altamirano:2016hug}.

This translates into linear behaviour of the scale factor as a function of comoving time, which in turn indicates a curvature dominated universe. Indeed, for $k=0$ there exists one positive real root of the characteristic equation \eqref{charac_eq} ($\omega_1$ shown in Fig.~\ref{fig:frequencies} with solid lines) while the other two are complex roots with real part negative (dashed lines). This means that for late conformal times we can write  
\beq
\texttt{a}^2(\tau)=A_1^{(0)}\exp(\omega_1\tau)\,, \label{exponential}
\eeq
which yields a scale factor in proper time
\beq
\texttt{a}(t)=\frac{\omega_1}{2}t\,, 
\label{physicaltime}
\eeq
where $\omega_1=0.793 (\gamma \hbar^2 c^4)^{1/3}$. By use of the Friedmann equation $H^2=\frac{\kappa}{3}\rho(t)$, where $H=\dot{\texttt{a}}/\texttt{a}$ (and the dot denotes the derivative with respect to comoving time $t$), we realize that $\rho(t)\propto \texttt{a}^{-2}$ which is equivalent to a universe dominated by a positive effective {\it{curvature like}} fluid ($K_{\text{eff}}=\omega_1^2/4$). From now on, when we refer to $K_{\text{eff}}$ we mean at the level of Friedmann equations --- the topology of spacelike surfaces does not change and is fixed by $k$.  This also implies that for late times the equation of state parameter $w$, defined as $P=w \rho$, is $w=-1/3$. One can see this with use of the second Friedmann equation $\dot{\rho}+3H (P+\rho)=0$ and the fact that $\rho(t)\propto \texttt{a}^{-2}$. We will make a deeper analysis of the equation of state in the next section. Note that because of the exponential behaviour of Eq.~\eqref{exponential} the expression \eqref{physicaltime} for the observed scale factor as a function of the observed proper time $t_p$ is the same i.e. $a_p=\frac{\omega_1}{2}t_p$ --- see ~\ref{ahamiltonian} for details.

\subsection{Emergent Dark Fluid}
\label{emfluid}

 For an classical observer unable to access the quantum nature of the scale factor, the universe will be described by Einstein equations
 \beq
 G_{ab}(\texttt{a}_p)=\kappa T_{ab}\,,
 \eeq
where $G_{ab}$ is the Einstein tensor and $T_{ab}$ is the energy momentum tensor. As noted above, because CCG does not break the homogenous and isotropic symmetries the effective metric for the observer is still of the form \eqref{new_metric} while the energy momentum tensor is the one associated with a perfect fluid 
\beq
T^{ab}=(\rho +P)U^aU^b+Pg^{ab}\,,
\label{emtccg}
\eeq
where $\rho$ is the energy density, $P$ the pressure,  $U^{a}=(\frac{\partial}{\partial t})^a$   the 4-velocity of the fluid and $g^{ab}$
 the metric. On the other hand, a perfect fluid can be characterized by its equation of state $w$ defined as $P=w\rho$. The Einstein equations guarantee the conservation of the energy momentum tensor $\nabla_aT^{ab}=0$ and reducing the the two Friedman equations
\begin{eqnarray}
H_u^2+\frac{k}{\texttt{a}^2}&=&\frac{\kappa}{3}\rho\,, \label{1friedeq}\\
\dot{\rho}-3 H\rho(1+w)&=&0\,, \label{eqstate}
\end{eqnarray}
where $H_u=\frac{\dot{\texttt{a}}}{\texttt{a}}$ is the  fiducial Hubble parameter.
The above equations can be solved to find 
\begin{eqnarray}
\rho(t)&=&\frac{3}{\kappa}\bigg(\frac{\dot{\texttt{a}}^2+k}{\texttt{a}^2}\bigg)\,, \label{rhoooo}\\
w(t)&=&-\frac{1}{3}\bigg(1+2\frac{\ddot{\texttt{a}}\texttt{a}}{\dot{\texttt{a}}^2+k}\bigg)\,. \label{wqqq}
\end{eqnarray}
Note that Eqs.~\eqref{1friedeq}-\eqref{wqqq} are for the fiducial quantities and the expression for the physical quantities can be read of Table~\ref{tquantities}. In particular we have that $w=w_p$ where $w_p$ is the physical observed equation of state. 
As previously advertised in the last section, for late times the scale factor  grows linearly with time, and so has $\ddot{\texttt{a}}=0$ and $\dot{\texttt{a}}=\alpha$ = constant, reducing the above equations to 
\beq
\rho=\frac{3}{\kappa}\frac{1}{\texttt{a}^2}(\alpha^2+k)\,,\,\,\,\,\,\,\,\,\,\,\,\, w=-\frac{1}{3}\,.
\eeq

\begin{figure*}
\minipage{0.32\textwidth}
  \includegraphics[width=\linewidth]{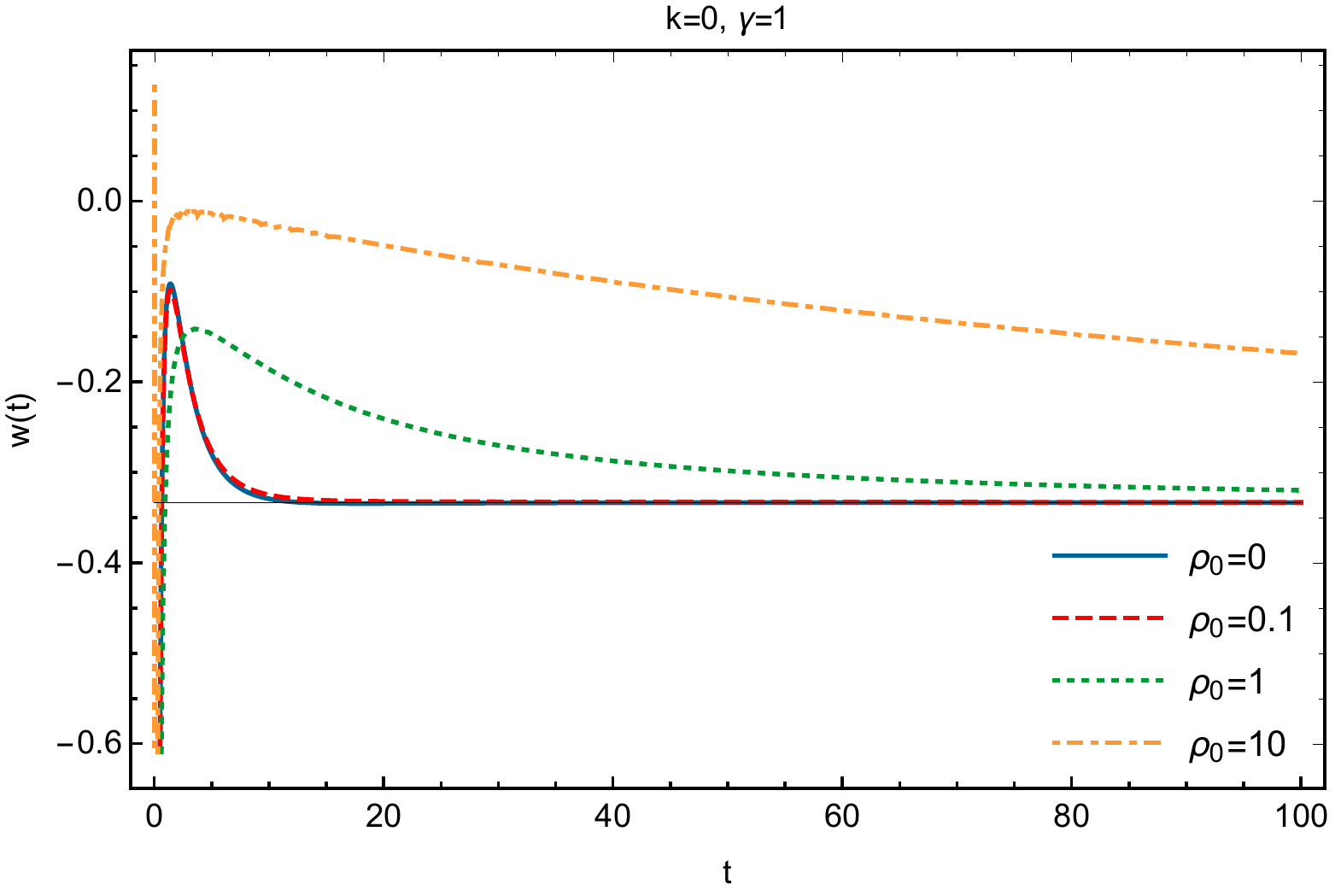}
  \endminipage\hfill
\minipage{0.32\textwidth}
  \includegraphics[width=\linewidth]{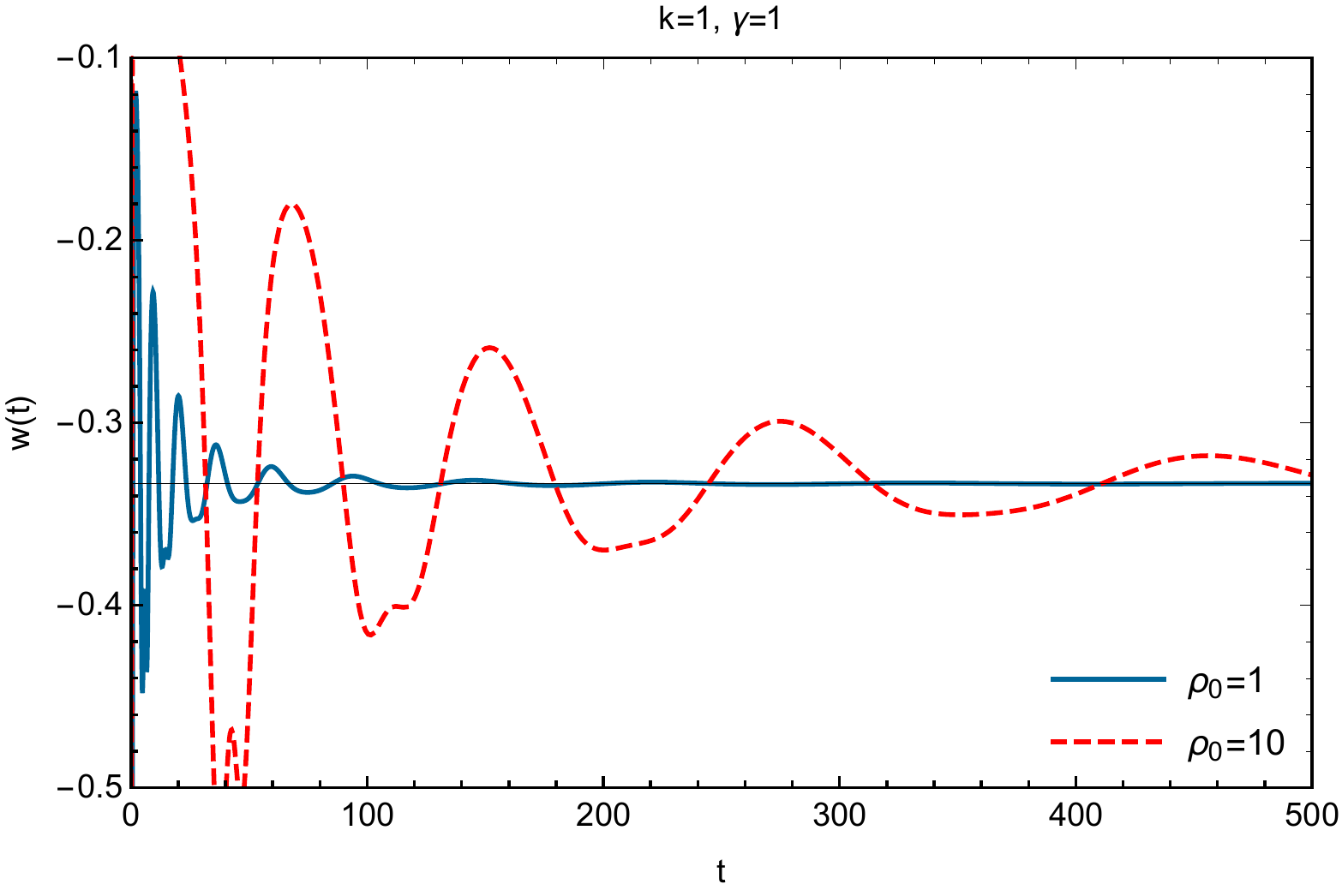}
  \endminipage\hfill
\minipage{0.32\textwidth}
  \includegraphics[width=\linewidth]{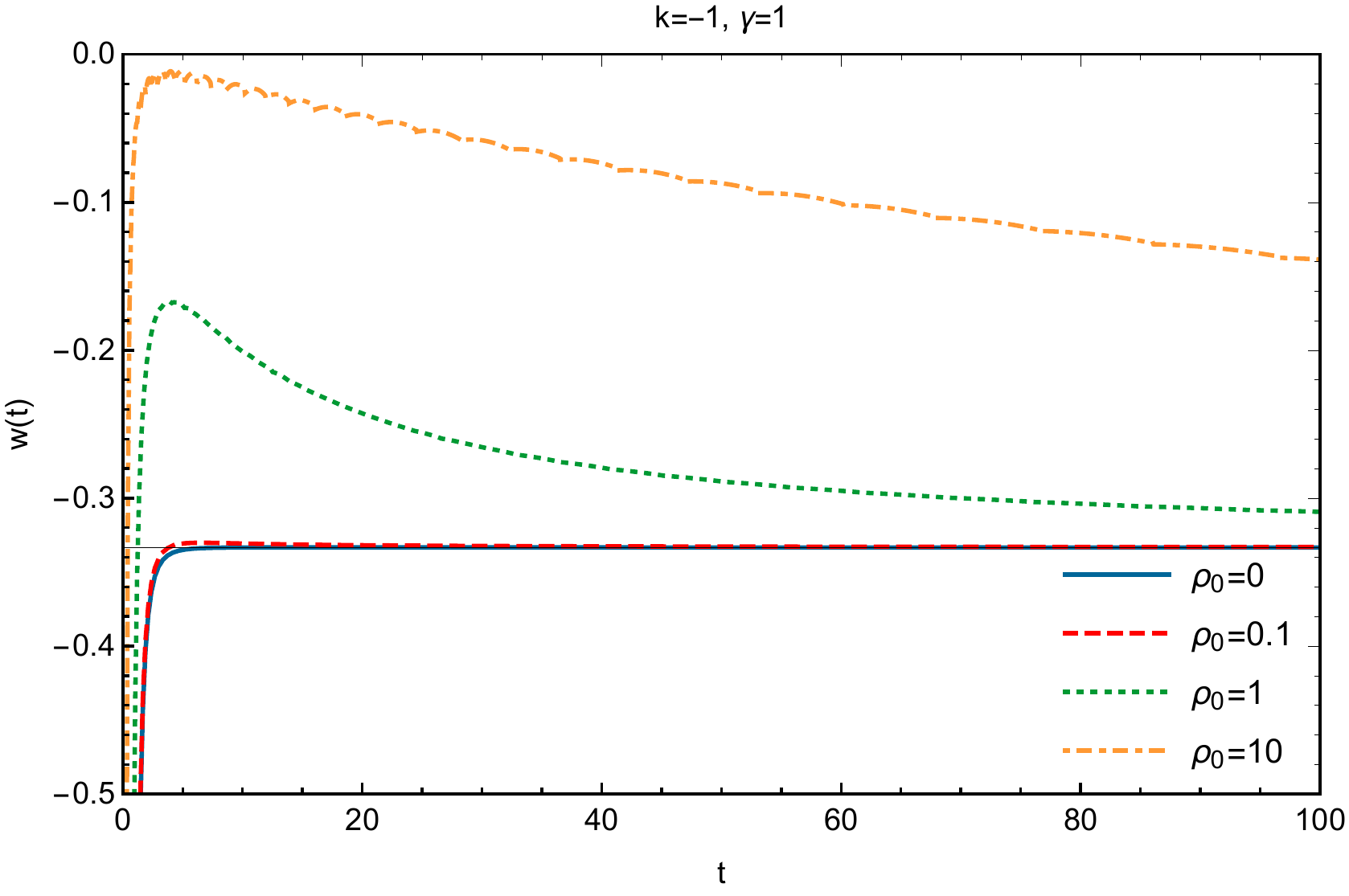}
  \endminipage\\
\minipage{0.32\textwidth}
  \includegraphics[width=\linewidth]{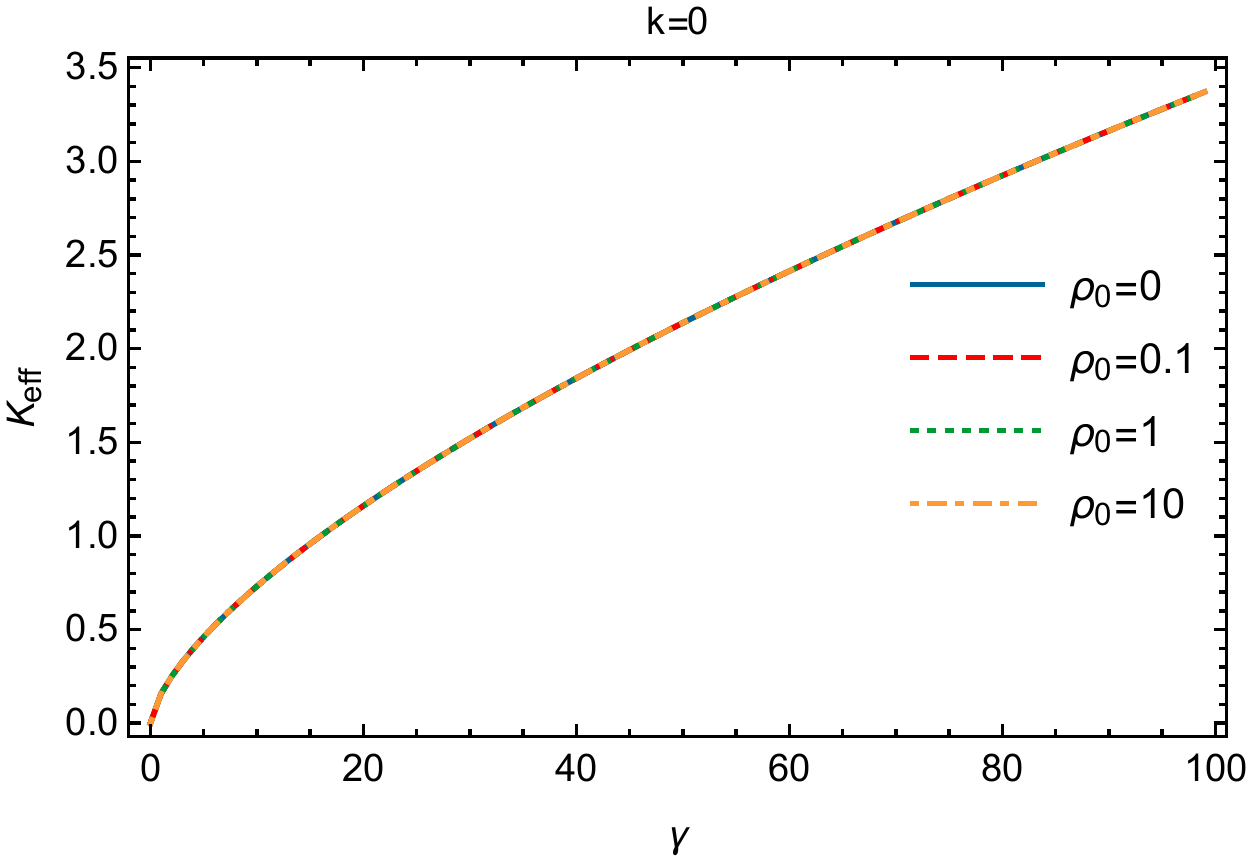}
  \endminipage\hfill
\minipage{0.32\textwidth}
  \includegraphics[width=\linewidth]{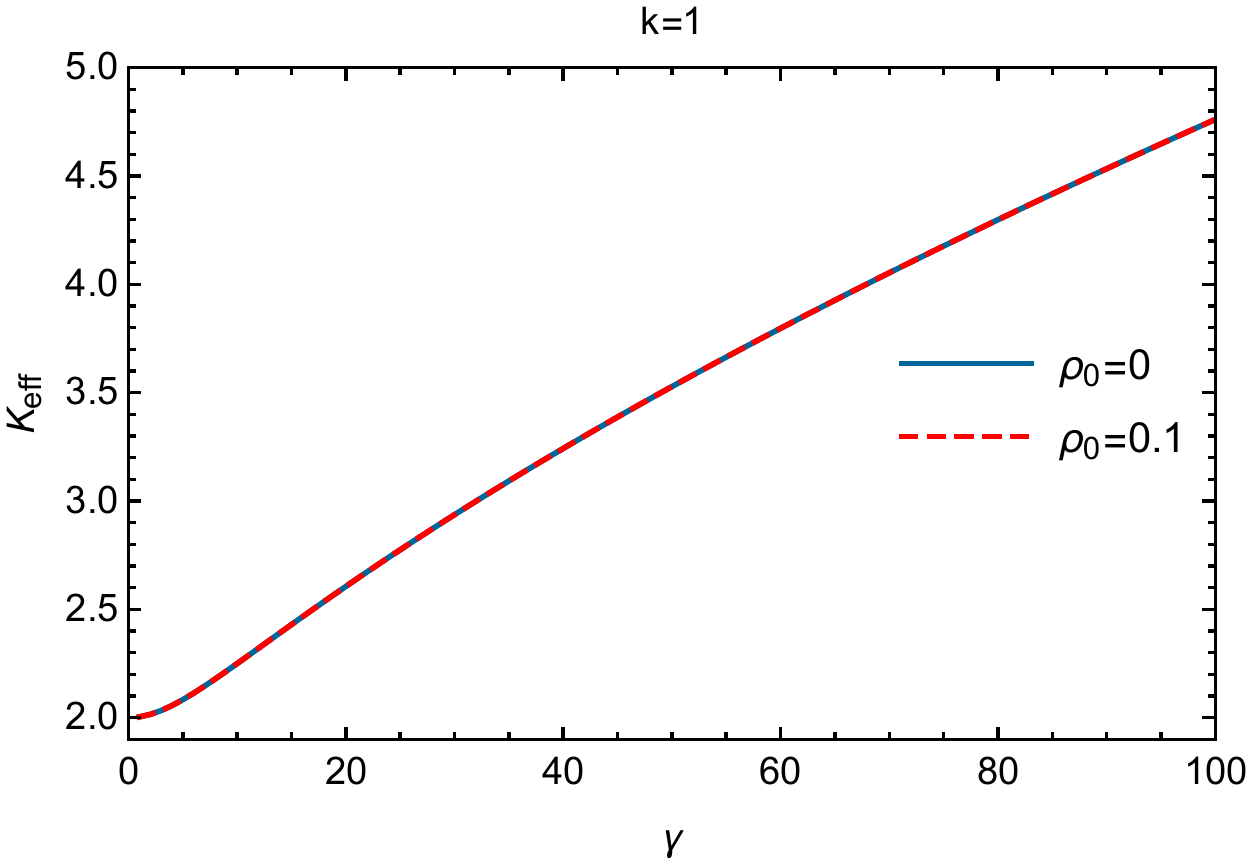}
  \endminipage\hfill
\minipage{0.32\textwidth}
  \includegraphics[width=\linewidth]{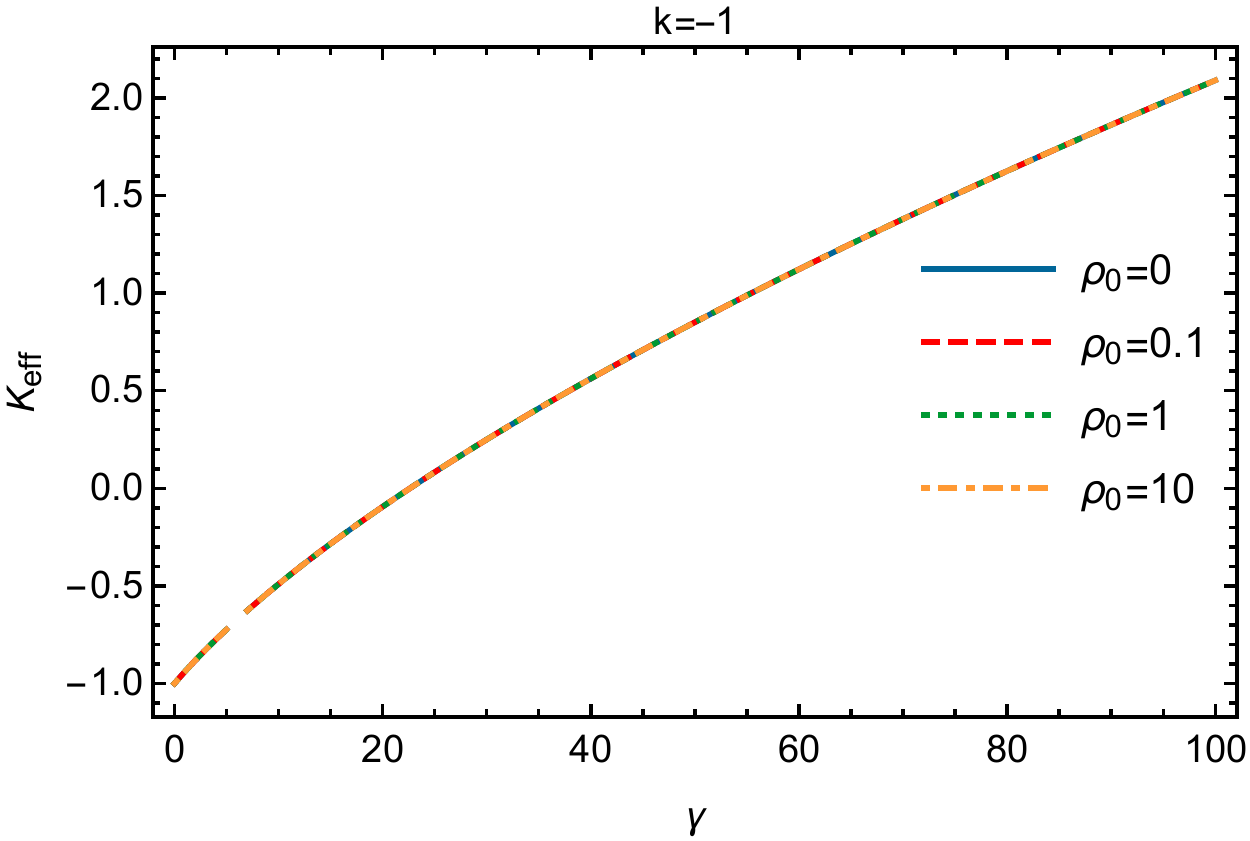}
  \endminipage
  \caption{Behaviour of the equation of state (top) and $K_{\text{eff}}$ (bottom) for the three different topologies. For large times the equation of state converges to the value $-1/3$ making the universe to be dominated by curvature. The value of the curvature parameter is shown in the bottom part and it does not depend on the value of $\rho_0$. This is because for late time the scale factor is dominated by the homogenous solution that is where the effect of CCG is much stronger than the initial matter content. The case $k=-1$ is the only case that admits a change in sign of $K_{eff}$, $K_{eff}=0$ (effectively flat - empty universe) can be fine tuned for $\gamma \sim 25$. }
 \label{fig:eqofstate}
\end{figure*}

We illustrate the behaviour of the equation of state as a function of comoving time for all values of $k$ and different values of $\rho_0$ in Fig.~\ref{fig:eqofstate} (first row) where we see the late time asymptote to $w=-1/3$ for all cases. This implies that the first Friedman equation \eqref{1friedeq} can be written as 
\beq
H_u^2-\frac{1}{\texttt{a}^2}(k+\alpha^2)=0\,,
\eeq
giving a universe dominated by curvature with an effective curvature constant $K_{\text{eff}}=2k+\alpha^2$. 

We  emphasize that this {\it{curvature domination}} is at the level of the Friedmann equations; $K_{\text{eff}}$ does not change the topology of the universe. For $k=0$ and $k=1$ the effective curvature is always positive. For $k=-1$ the universe could be hyperbolic if $\alpha^2< 2$. In the second row of Fig.~\ref{fig:eqofstate} we present $K_{\text{eff}}$ for the three different values of $k$ as a function of $\gamma$. As expected, $K_{\text{eff}}$ is insensitive to the value of $\rho_0$ and it is consistent with the analysis that at late times the homogenous part of the solution of the scale factor dominates. 

\subsubsection{Separation}

As outlined  in Sec.~\ref{emfluid} the classical emergent fluid is composed of a purely emergent part due to CCG and a part due to the primordial dust $T^{ab}_\texttt{d}$.  Here we   outline the strategy for analyzing the classical perfect fluid $T^{ab}$ and how the primordial dust $T^{ab}_\texttt{d}$ affects it. 

To this end, note that in CCG the comoving time $t$ is defined {\it{a posteriori}} once we have solved Eq.~\eqref{master2} for the scale factor $\texttt{a}$. Also note that  $t$ is in general different from  the proper time  $t_\texttt{d}$ of the primordial dust. These two coordinates are respectively defined as 
       \beq
   \texttt{a}(\tau)d\tau=dt\,,\,\,\,\,\,\,\,\,\,\,a(\tau)d\tau=dt_\texttt{d}\,,
   \eeq
    where $a(\tau)$ is the scale factor entering in Eq.~\eqref{dust_scale} and $\texttt{a}(\tau)$ is the solution of Eq.~\eqref{classicala_def}.  With these definitions we can write the energy-momentum tensor of the dust defined by Eq.~\eqref{dustT} in terms of the classical scale factor $\texttt{a}$ and its associated comoving time 
    \beq
    T^{ab}_\texttt{d}=\rho \,U^{a}_\texttt{d}U^{b}_\texttt{d}= \rho\, \frac{\texttt{a}^2}{a^2}\,U^a U^b = \rho_N \,U^a U^b  \,\,\,\,\,\,\,\, U^a=\bigg(\frac{\partial}{\partial t}\bigg)^a \,,
    \label{dust_Tmunu}
    \eeq 
  where $\rho_N=\rho\, \frac{\texttt{a}^2}{a^2}$  and $U^a$ is the four-velocity of the classical fluid $T^{ab}$. We are now interested in whether the dust energy-momentum tensor defined with the previous equation is conserved or not. The divergence of  $T^{ab}_\texttt{d}$ in terms of the covariant derivative associated with the metric \eqref{new_metric} yields
 \beq  \label{conservation_new}
 \nabla_{a}T^{ab}_\texttt{d} =    \frac{d\rho_N}{dt}-3H_u\rho_N=\rho \frac{\texttt{a}}{a^2}\bigg(\frac{d\,a}{d\,t_\texttt{d}}-\frac{d\texttt{a}}{dt}\bigg) \,.
    \eeq  
     The right-hand side of   \eqref{conservation_new}
is generically non-zero. Hence the original dust fluid $T^{ab}_\texttt{d}$ will not be conserved in the classical observed spacetime.

In other words, CCG implies rather naturally a perfect fluid {\it{emergent}} from Eq.~\eqref{master2} that we denote by $T^{ab}$ (since it is the perfect fluid a classical observer will detect) and that is conserved under the covariant derivative associated with the metric \eqref{new_metric}. This emergent fluid has a component that is intrinsic to CCG and another contribution from the primordial dust $T^{ab}_\texttt{d}$. We have shown that we cannot treat $T^{ab}_\texttt{d}$ as a conserved disjoint fluid but we can still describe its properties using
    \beq
    T^{ab}=T^{ab}_D+T^{ab}_\texttt{d}\,,
    \eeq
 where $T^{ab}$ was studied in the previous section and $T^{ab}_\texttt{d}$ is defined by equation \eqref{dust_Tmunu}. By use of this equation we can write the relation between energy densities and pressures 
 \beq
 \rho=\rho_D+\rho_{o}\,,\,\,\,\,\,\,\,\, P=P_D\,,
 \eeq
 where $\rho_o$ denotes the  energy density of the primordial dust in the new coordinates \eqref{dust_Tmunu}
 \beq
 \rho_o=\rho_0 \frac{\texttt{a}^2}{a^5}\,.
 \eeq
 In the last expression all the quantities are functions of the new comoving time $t$. Note that  $a(\tau)$ as a function of conformal time is the solution to a Friedman universe filled with dust and thus 
 \beq
 a(\tau)=\frac{\kappa}{3}\rho_0\left\{
  \begin{array}{@{}ll@{}}
   \sin(\tau/2)^2, & k=1 \\
    \tau^2/4, & k=0\\
   \sinh(\tau/2)^2, & k=-1
  \end{array}\right.
  \eeq

Since an observer in the universe will only infer from measurement one conserved fluid, he/she will not be able to distinguish between the parts ($\rho_\texttt{d}$ , $\rho_{o}$) that form it. Nevertheless, as shown in Fig.~\ref{fig:scale} the presence of $\rho_o$ imparts a quantitative difference to the evolution of the universe as compared to the $\rho_o=0$ case.

% 
% One could also define a equation of state associated to all the quantities intrinsic to CCG
% \beq
%P_\texttt{d} =\rho_\texttt{d}\omega_\texttt{d}\,,
%\eeq
%that can be rewritten as 
%\beq
%\omega_\texttt{d}=\omega\frac{\rho}{\rho_\texttt{d}}\,.
%\eeq
% 
\section{Discussions} \label{secdis}
Perhaps, the most striking feature of  our CCG model with matter is the fact that if the primordial dust is treated as a classical fluid, its influence in the universe will be {\it{washed out}} for late times -- where the decoherence (dominated by $\gamma$) takes over. An observer at late times will be unable to discern whether or not the universe had some primordial classical matter.  On the other hand, the primordial fluid does not have the same consequences as dust has in classical General Relativity. As seen in Fig.~\ref{fig:frequencies} (top), the primordial dust makes the universe expand faster, contrary to the intuition one has from GR -- the dust makes the universe contract. In CCG there is an interplay between the value of the primordial dust and the decoherence parameter $\gamma$ conspiring to produce such an effect. Note, however, that in the limit where $\gamma=0$ (which in turns correspond to a fully classical description) we recover the Friedman equations with dust. Nonetheless, this is not fully satisfactory, since Friedman evolution has been very well constrained by data \cite{Ade:2015xua}, questioning the necessity to introduce CCG in the first place. 

This leaves open the question of how to include matter in the model. Phenomenologically, one could argue that the observed energy momentum tensor  \eqref{emtccg} is indeed composed of cold dark matter and a cosmological constant plus a CCG contribution. The first Friedman equation for the values today can be written as 
\beq
1=\Omega_{b}^0+\Omega_{\text{cdm}}^0+\Omega_{\text{rad}}^0+\Omega_{\Lambda}^0+\Omega_{k}^0+\Omega_{\text{CCG}}^0\,,
\eeq
where $\Omega_{i}^0=\frac{\rho_i(t_0)}{\rho_c^0}$ for the respective baryon, cold dark matter, radiation, vacuum, curvature, and CCG
contributions, with $\rho_c^0$ the value of the critical density today. Using the results  for the cosmological parameters given by Planck 
\cite{Ade:2015xua}
we can constrain
\beq
0\leq\Omega_{k}^0+\Omega_{\text{CCG}}^0\leq 0.001\pm0.004\,, \label{planckconst}
\eeq
In an universe where $k=0$,  since $\texttt{a}_p=\frac{\omega_1}{2}t_p$ for late times, this implies (upon use of Friedmann equation) that $\rho_{\text{CCG}}=3 \omega_1^2/(4\kappa \texttt{a}_p^2)$, where $\omega_1=0.7
(\hbar^2 c^4\gamma)^{1/3}$. With this we can write Eq.~\eqref{planckconst} as\beq
\Omega_{ccg}=\frac{\omega_1^2}{4 (\texttt{a}_p^0)^2(H_p^0)^2}<0.001\,,
\eeq
where the superscript $0$ denotes the values today. We finally find
\beq
\gamma_p<0.1(\texttt{a}_p^0H_p^0)^3\bigg(\frac{V_0}{\kappa \hbar c^2}\bigg)^2\sim \frac{10^{66}}{m^6 s}V_0^2\,.
\eeq
Taking the fiducial volume to be the order of the Hubble volume, one finds that $\gamma_p<10^{266}$!!. This implies that present cosmological observations are far from ruling out this model. Conversely, one could also conclude that the cosmological effects of the CCG model  are of  negligible observational consequence.

Is this a game stopper? As discussed above there is still work to be done in trying to understand how to include the presence of classical and quantum matter in CCG in a way that i) is able to reproduce the strongly constrained behaviour of the $\Lambda$CDM model and ii) gives a possible resolution to the unsolved problems of this standard model of cosmology.

%suggesting than $\gamma<$. Of course, this upper bound is a very crude overestimation, and a model that is able to introduce at the fundamental level the observed matter and CCG will definitely better constrain $\gamma$.

\section{Conclusion} \label{secconc}

We have  studied the cosmological consequences of the CCG model in a universe containing primordial dust.  We found that for late times the homogenous solution dominates, meaning that the fundamental CCG modification is stronger than that of a primordial perfect fluid. Extending the analysis in \cite{Altamirano:2016hug}, we have furthermore found that the equation of state asymptotes to $w=-1/3$, yielding a universe that is curvature-dominated, and computed the effective `curvature fluid'. This shows that the emergent dark fluid behaves as a curvature parameter in the Einstein equations at late times.

More realistic  extensions of this model would include incorporating both vacuum energy and primordial radiation, which should correspond to the effective stress energy of the elementary particles of the standard model.  More generally one could add
  a conserved energy momentum tensor representing  different matter components of the universe, analyzing the effect that each has  on the evolution of the scale factor in the CCG context. Constrained by current observations \cite{Ade:2015xua}, the curvature of the observable universe is close to zero suggesting that CCG modifications to Friedman equations must be small enough to i) predict a 
vacuum-dominated universe and  ii) give $\Omega_k \sim 0$. 
Finally, it would be interesting to study CCG in the context of theories of gravity that violate energy conservation to better understand the observable consequences of CCG and construct a comparison background of theories of modified gravity. 

CCG is rooted in the idea that gravitational screening is not possible, and so implies that spacetime has an irreducible amount
of `noisy' interaction with any form of stress-energy. In the cosmological context we restrict this situation to that of a test particle  {\it{interacting}} with a classical potential (metric) that is sourced by fundamentally quantized degrees of freedom. This modification of quantum mechanics has been extensively studied in a variety of different situations \cite{1742-6596-306-1-012006,PhysRevD.93.024026,PhysRevLett.80.2508,PhysRevLett.93.240401,PhysRevD.14.2460,PhysRevD.52.2176,Altamirano:2016knc,2016arXiv160504302G},  with a recent particular application to cosmology in \cite{Altamirano:2016hug,Josset:2016vrq}. Whereas previous attempts to study modifications of quantum mechanics in the cosmological context have imposed this modification  by hand, CCG attempts to construct from first principles a modified Wheeler deWitt equation. 

\section*{Acknowledgements}
RP would like to thank the hospitality of the University of Waterloo where this project began.  Research at Perimeter Institute is supported by the Government of Canada through Industry Canada and by the Province of Ontario through the Ministry of Research and Innovation. This work was supported in part by the Natural Science and Engineering Research Council of Canada.

%
%
%\naty{COmments about how to treat the matter present and give a discussion of the full quantum matter there and how CCG can help to resolve the problem of gravitational backreaction with quantum matter\\
%comment on possible ways to actually see real matter in the universe like dust radiation, etc....\\
%comments on possible testability of our model: like graviataional waves, pulsar timing.\\
%extentions of these mode with Feedback? \\
%Relation of the derivative of the hamiltonian with the hamilotonian of the emergent fluid?\\}
%%\bibliographystyle{unsrt}

%Passage au mode appendice
%\renewcommand{\thesection}{\Alph{section}} %On change la numérotation pour avoir des lettres
%
%\vfill\eject  
%\newpage

\appendix

\section{Notation} \label{app1}
Through this work we used different notations to indicate the difference between underlaying physical quantities, observed physical quantities and fiducial quantities (that were defined after physical quantities for convenience):\\

Physical quantities: $a_p, p_p,\rho_{0p}$.\\

Observed quantities: $\texttt{a}_p, \gamma_p, t_p,  H_p, \rho_{\text{CCG}},  w_p,\Omega_{\text{CCG}}$.\\

Fiducial quantities: $a,p, \rho_0,\hat{a}, \hat{p}, \texttt{a}, t, \gamma,H_u,\rho,w$.\\

The relation between the fiducial quantities and the physical and observed quantities is summarized in Table~\ref{tquantities}.

%\bwt

\begin{table}
\caption{\label{tquantities}
Relation between physical, observed and fiducial quantities where $\eta=\sqrt{\frac{\kappa}{3 V_0}}$. Note that the time derivatives denoted by an upper dot are with respect of $t_p$/$t$ if applied to physical/fiducial quantities.}

\noindent \centering{}

\begin{tabular}{|c |c| c| c|}
\hline \hline
Name & Description & Definition & Properties\\ \hline \hline
$a_p$ & underlying  scale factor &&\\
$p_p$ & canonical conjugate momentum && $\{a_p,p_p\}=1$\\
$\h a_p$ & scale factor operator &&\\
$\h \pi$ & momentum density operator &&\\
$\h p_p$ & physical momentum operator &$p=\int dV \pi$& $[\h a_p,\h p_p]=i\hbar$\\
$\gamma_p$&physical decoherence parameter& See~\eqref{physicalmaster}&\\
$\rho_{0p}$& primordial dust& &\\ \hline \hline
$\texttt{a}_p$& observed scale factor & $\sqrt{\mean{\h a_p^2}}$& \\
$t_p$& observed comoving time & $\texttt{a}_pd\tau=dt_p$&\\
$H_p$& observed Hubble parameter & $\frac{\dot{\texttt{a}}_p}{\texttt{a}_p}$& $H_p^2+\frac{k}{\texttt{a}_p^2}=\frac{\kappa}{3}\rho_{\text{ccg}}$\\
$\rho_{\text{ccg}}$& observed dark energy density&& $\rho_{\text{ccg}}=\frac{3}{\kappa}(\frac{\dot{\texttt{a}}_p^2+k}{\texttt{a}_p^2})$\\
$w_p$& observed equation of state& $P_{\text{ccg}}=w_p\rho_{\text{ccg}}$&$w_p=\frac{\dot{\rho}_{\text{ccg}}}{3H_p\rho_{\text{ccg}}}-1$\\
$\Omega_{\text{ccg}}$&&$\frac{\rho_{\text{ccg}}\kappa}{3H_p^2}$&\\ \hline \hline
$a$&fiducial scale factor&$a_p/\eta$&\\
$p$&fiducial momentum&$\eta\, p_p$&$\{a,p\}=1$\\
$\rho_0$&fiducial primordial dust&$\rho_{0p}\,\eta\,V_0$&\\
$\h a$&fiducial scale factor operator&$\h a_p/\eta$&\\
$\h p$&fiducial momentum operator&$\h p_p\,\eta$&$[\h a,\h p]=i\hbar$\\
$\gamma$&fiducial decoherence parameter&$\gamma_p\,\eta^4$&See \eqref{truemaster}\\
$\texttt{a}$&fiducial observed scale factor&$\texttt{a}_p/\eta$&$\sqrt{\mean{\h a^2}}$\\
$t$&fiducial comoving time&$t_p/\eta$&$\texttt{a}d\tau=dt$\\
$H_u$&fiducial Hubble parameter&$\eta\, H_p$&$\frac{\dot{\texttt{a}}}{\texttt{a}}$\\
$\rho$&fiducial dark energy density&$\eta^2\rho_{\text{ccg}}$&$\rho=\frac{3}{\kappa}(\frac{\dot{\texttt{a}}^2+k}{\texttt{a}^2})$\\
$w$&fiducial equation of state&$w=w_p$&$w_p=\frac{\dot{\rho}}{3H\rho}-1$\\ \hline

\hline

\end{tabular}
\end{table}

%\ewt

\section{Physical quantities and Hamiltonian density } \label{ahamiltonian}

From the action of GR with a fluid 
\beq
S=\int d^4x\sqrt{-g}\bigg[\frac{R}{2\kappa}-\rho_p\bigg]\,,
\eeq
 and the metric 
 \beq
 ds^2=a_p^2[n^2(\tau)d\tau^2+\frac{dr^2}{\sqrt{1-kr^2}}+r^2d\theta^2+r^2\sin^2(\theta)d\phi^2]\,,
 \eeq
  we can compute the Lagrangian density 
 \beq
 {\cal{L}}=\frac{3 a_p^2n(\tau)k}{\kappa}-\frac{3a_p'^2}{n(\tau)\kappa c^2}-\rho_p a_p^4n(\tau)\,,
 \eeq
 where we have denoted $a_p$ and $\rho_p$ the physical scale factor and physical matter and $a_p'$ denotes derivative with respect to the conformal time. In order to compute the Hamiltonian density we define the momentum density as
 \beq
 \pi=\frac{\partial {\cal{L}}}{\partial a_p'}=-\frac{6 a_p'}{n(\tau)\kappa c^2}\,,
 \eeq
 and we obtain
 \beq
 {\cal{H}}=\pi\,a_p'-{\cal{L}}=-\frac{\pi^2\kappa c^2}{12}-3\frac{a_p^2k}{\kappa}-\rho_p a_p^4\,.
 \eeq
 Upon defining the physical momentum $p_p=\int \pi d^3x $ we can compute the Hamiltonian for the system
 \beq
 H=-\frac{p_p^2 \kappa c^2}{12 V_0}-3\frac{a_p^2 k V_0}{\kappa}-\rho_p a_p V_0\,,
 \eeq
 where $V_0=\int d^3 x$ is a fiducial volume which in the coordinates we have chosen is $V_0=\int \frac{r^2}{\sqrt{1-k r^2}}\sin(\theta)dr d\theta d\phi$. 
 To obtain the Hamiltonian \eqref{hamil} we rescale the scale factor and the momentum in the following way
 \beq
 a=\sqrt{\frac{3 V_0}{\kappa}}a_p\,,\,\,\,\,\,\,\,\,\,\,\,\,\,\,\, p=\sqrt{\frac{\kappa}{3 V_0}}p_p\,. \label{relationscale}
 \eeq
 Note that with this definitions $a$ and $p$ are still canonical conjugate variables since their Poisson bracket is equal to one. Finally, we have also rescaled $\rho_{0}=\rho_{0p}\sqrt{\kappa\,V_0 /3}$ to find
 \beq
 H=-\frac{p^2c^2}{4}-k a^2+\rho_0 a\,.
 \eeq
 
 \subsection{Quantization and  Master equation} \label{amaster}
 
 Since the variables $a$ and $p$ are canonically conjugate their conmutator is $[a,p]=i\hbar$. However, 
 the modification of the Wheeler deWitt occurs in physical space 
 \beq
  \frac{d\hat{\varrho}}{d\tau} = -\frac{i}{\hbar} [\hat{H}(a_p,p_p), \hat{\varrho} ] - \frac{\gamma_p}{8} [\hat{a}_p^2,[\hat{a}_p^2, \hat{\varrho} ] ]\,,
  \label{physicalmaster}
 \eeq
where $\gamma_p$ is the decoherence parameter which has units of $1/T$. Upon normalization of the quantities we can write the master equation as
\beq
 \frac{d\hat{\varrho}}{d\tau} = -\frac{i}{\hbar} [\hat{H}(a,p), \hat{\varrho} ] - \frac{\gamma}{8} [\hat{a}^2,[\hat{a}^2, \hat{\varrho} ] ]\,, \label{truemaster}
 \eeq
 where $\gamma=\gamma_p\frac{\kappa^2}{9 V_0^2}$\,. Eq.~\eqref{truemaster} is the master equation we employ in this paper. \\

\subsection{Observed physical quantities}
\label{obsphys}
As discussed in the introduction the observed scale factor of an observer unaware of the underlying quantum mechanism will be
\beq
\texttt{a}_p\equiv\sqrt{\mean{\hat{a}_p^2}}\,.
\eeq
 On the other hand we will compute all quantities and show plots of quantities involving $a$ which is related with $a_p$ via 
 Eq.~\eqref{relationscale}. This implies that is
 \beq
 \texttt{a} \equiv \sqrt{\mean{\hat{a}^2}}=\sqrt{\frac{3 V_0}{\kappa}}\texttt{a}_p\,.
 \eeq
Upon solving the system of equations \eqref{eqc1}--\eqref{eqc3} we will get an expression for $\texttt{a}(\tau)$ and we will use the relation $a(\tau)d\tau=dt$ to find the functional form of $a(t)$. With this relations we also find that the observed proper time is 
 \beq
 t_p\equiv \int \texttt{a}_p(\tau) d\tau=\sqrt{\frac{\kappa}{3 V_0}} t\,.
 \eeq
 Note that in the case where $\texttt{a}(\tau)\propto \exp( \omega \tau)$ (we find this behaviour for late times in CCG regardless of the value of $k$) we have $\texttt{a}(t)=\texttt{a}_p(t_p)$ with no explicit dependence in $V_0$ of the observed scale factor. Nonetheless, the time $\bar{t}$ after which the behaviour of $\texttt{a}$ is well approximated by a linear function implies that the observed scale factor is well approximated by the same linear function only for $t_p>\bar{t}_p=\sqrt{\frac{\kappa}{3 V_0}} \bar{t}$.

   \section*{References}
   \bibliographystyle{iopart-num}

\bibliography{biblio}

\providecommand{\newblock}{}
\begin{thebibliography}{10}
\expandafter\ifx\csname url\endcsname\relax
  \def\url#1{{\tt #1}}\fi
\expandafter\ifx\csname urlprefix\endcsname\relax\def\urlprefix{URL }\fi
\providecommand{\eprint}[2][]{\url{#2}}

\bibitem{PhysRevA.63.022101}
Peres A and Terno D~R 2001 {\em
  \href{http://link.aps.org/doi/10.1103/PhysRevA.63.022101}{Phys. Rev. A}\/}
  {\bf 63}(2) 022101

\bibitem{carlip_is_2008}
Carlip S 2008 {\em
  \href{http://stacks.iop.org/0264-9381/25/i=15/a=154010}{Class. Quantum
  Grav.}\/} {\bf 25} 154010 ISSN 0264-9381

\bibitem{albers_measurement_2008}
Albers M, Kiefer C and Reginatto M 2008 {\em
  \href{http://link.aps.org/doi/10.1103/PhysRevD.78.064051}{Phys. Rev. D}\/}
  {\bf 78} 064051

\bibitem{2014NJPh...16f5020K}
Kafri D, Taylor J~M and Milburn G~J 2014 {\em
  \href{http://stacks.iop.org/1367-2630/16/i=6/a=065020}{New Journal of
  Physics}\/} {\bf 16} 065020

\bibitem{aharonov_how_1988}
Aharonov Y, Albert D~Z and Vaidman L 1988 {\em
  \href{http://link.aps.org/doi/10.1103/PhysRevLett.60.1351}{Phys. Rev.
  Lett.}\/} {\bf 60} 1351--1354

\bibitem{diosi_models_1989}
Di{\'o}si L 1989 {\em
  \href{http://link.aps.org/doi/10.1103/PhysRevA.40.1165}{Phys. Rev. A}\/} {\bf
  40} 1165--1174

\bibitem{penrose_gravitys_1996}
Penrose R 1996 {\em
  \href{http://link.springer.com/article/10.1007/BF02105068}{Gen Relat
  Gravit}\/} {\bf 28} 581--600 ISSN 0001-7701, 1572-9532

\bibitem{Altamirano:2016fas}
Altamirano N, Corona-Ugalde P, Mann R~B and Zych M 2016  (\textit{Preprint}
  \eprint{https://arxiv.org/abs/1612.07735})

\bibitem{Altamirano:2016hug}
Altamirano N, Corona-Ugalde P, Khosla K, Mann R~B and Milburn G 2017 {\em
  {\href{http://iopscience.iop.org/article/10.1088/1361-6382/aa6d41/meta}{Class.
  Quant. Grav.}}\/} {\bf 34} 115007 (\textit{Preprint}
  \eprint{https://arxiv.org/abs/1605.05980})

\bibitem{diosi_quantum_2000}
Di{\'o}si L, Gisin N and Strunz W~T 2000 {\em
  \href{http://link.aps.org/doi/10.1103/PhysRevA.61.022108}{Phys. Rev. A}\/}
  {\bf 61} 022108

\bibitem{diosi_quantum_1995}
Diosi L 1995 {\em arXiv:quant-ph/9503023\/}
  \urlprefix\url{http://arxiv.org/abs/quant-ph/9503023}

\bibitem{Yang:2017xyh}
Yang I~S 2017  (\textit{Preprint} \eprint{https://arxiv.org/abs/1703.03466})

\bibitem{1742-6596-306-1-012006}
Diósi L 2011 {\em
  \href{http://stacks.iop.org/1742-6596/306/i=1/a=012006}{Journal of Physics:
  Conference Series}\/} {\bf 306} 012006

\bibitem{PhysRevD.93.024026}
Tilloy A and Di\'osi L 2016 {\em
  \href{https://link.aps.org/doi/10.1103/PhysRevD.93.024026}{Phys. Rev. D}\/}
  {\bf 93}(2) 024026

\bibitem{PhysRevLett.80.2508}
Garay L~J 1998 {\em
  \href{https://link.aps.org/doi/10.1103/PhysRevLett.80.2508}{Phys. Rev.
  Lett.}\/} {\bf 80}(12) 2508--2511

\bibitem{PhysRevLett.93.240401}
Gambini R, Porto R~A and Pullin J 2004 {\em
  \href{https://link.aps.org/doi/10.1103/PhysRevLett.93.24040}{Phys. Rev.
  Lett.}\/} {\bf 93}(24) 240401

\bibitem{PhysRevD.14.2460}
Hawking S~W 1976 {\em
  \href{https://link.aps.org/doi/10.1103/PhysRevD.14.2460}{Phys. Rev. D}\/}
  {\bf 14}(10) 2460--2473

\bibitem{PhysRevD.52.2176}
Unruh W~G and Wald R~M 1995 {\em
  \href{https://link.aps.org/doi/10.1103/PhysRevD.52.2176}{Phys. Rev. D}\/}
  {\bf 52}(4) 2176--2182

\bibitem{Altamirano:2016knc}
Altamirano N, Corona-Ugalde P, Mann R~B and Zych M 2017 {\em
  {\href{http://iopscience.iop.org/article/10.1088/1367-2630/aa551b/meta}{New
  J. Phys.}}\/} {\bf 19} 013035 (\textit{Preprint}
  \eprint{https://arxiv.org/abs/1605.04312})

\bibitem{2016arXiv160504302G}
Grimmer D, Layden D, Mann R~B and Mart\'{\i}n-Mart\'{\i}nez E 2016 {\em
  {\href{http://link.aps.org/doi/10.1103/PhysRevA.94.032126}{Phys. Rev. A}}\/}
  {\bf 94}(3) 032126

\bibitem{bassi_models_2013}
Bassi A, Lochan K, Satin S, Singh T~P and Ulbricht H 2013 {\em
  {\href{http://link.aps.org/doi/10.1103/RevModPhys.85.471}{Rev. Mod.
  Phys.}}\/} {\bf 85} 471--527

\bibitem{Josset:2016vrq}
Josset T, Perez A and Sudarsky D 2017 {\em
  {\href{https://journals.aps.org/prl/abstract/10.1103/PhysRevLett.118.021102}{Phys.
  Rev. Lett.}}\/} {\bf 118} 021102 (\textit{Preprint}
  \eprint{https://arxiv.org/abs/1604.04183})

\bibitem{Kafri:2015iha}
Kafri D, Milburn G~J and Taylor J~M 2015 {\em
  {\href{http://iopscience.iop.org/article/10.1088/1367-2630/17/1/015006}{New
  J. Phys.}}\/} {\bf 17} 015006

\bibitem{Khosla:2016tss}
Khosla K and Altamirano N 2017 {\em
  {\href{https://link.aps.org/doi/10.1103/PhysRevA.95.052116}{Phys. Rev.}}\/}
  {\bf A95} 052116 (\textit{Preprint}
  \eprint{https://arxiv.org/abs/1611.09919})

\bibitem{PhysRevA.36.5543}
Caves C~M and Milburn G~J 1987 {\em
  {\href{http://link.aps.org/doi/10.1103/PhysRevA.36.5543}{Phys. Rev. A}}\/}
  {\bf 36}(12) 5543--5555

\bibitem{Padilla:2014yea}
Padilla A and Saltas I~D 2015 {\em Eur. Phys. J.\/} {\bf C75} 561
  (\textit{Preprint} \eprint{https://arxiv.org/abs/1409.3573})

\bibitem{Ade:2015xua}
Ade P~A~R {\em et~al.\/} (Planck) 2016 {\em Astron. Astrophys.\/} {\bf 594} A13
  (\textit{Preprint} \eprint{https://arxiv.org/abs/1502.01589})

\end{thebibliography}

\end{document}